\documentclass[a4paper,11pt]{article}
\usepackage{jinstpub} 
\usepackage{siunitx}
\usepackage{lipsum}
\usepackage{graphicx}
\usepackage{caption}
\usepackage{subcaption}
\usepackage{booktabs}
\usepackage{multirow}
\usepackage{rotating}
\usepackage{xfrac}
\usepackage{nicefrac}
\usepackage[normalem]{ulem}
\usepackage{orcidlink}

\usepackage[section]{placeins} 

\newcommand{\Autoref}[1]{%
	\begingroup%
	\def\chapterautorefname{Chapter}%
	\def\sectionautorefname{Section}%
	\def\subsectionautorefname{Subsection}%
	\autoref{#1}%
	\endgroup%
}

\newcommand{\model}{\text{R12699-406-M4}~}
\newcommand{\modelshort}[1]{\text{R12699}}

\newcommand{\ca}{{\sim}}

\newlength{\twosubht}
\newsavebox{\twosubbox}

\DeclareSIUnit\voltpp{Vpp}

\title{Characterization of the Hamamatsu R12699-406-M4 Photomultiplier Tube in Cold Xenon Environments}

\author[a]{M.~Adrover\orcidlink{0009-0009-4903-4125},}
\author[a]{L.~Baudis\orcidlink{0000-0003-4710-1768},}
\author[a]{A.~Bismark\orcidlink{0000-0002-0574-4303},}
\author[b]{A.~P.~Colijn\orcidlink{0000-0002-3118-5197},}
\author[a]{J.~J.~Cuenca-García\orcidlink{0000-0002-3869-7398},}
\author[b]{M.~P.~Decowski\orcidlink{0000-0002-1577-6229},}
\author[b]{M.~Flierman\orcidlink{0000-0002-3785-7871},}
\author[b]{T.~den Hollander\orcidlink{0009-0004-5456-5995}}
\affiliation[a]{Physik-Institut, University of Z\"urich, 8057  Z\"urich, Switzerland}
\affiliation[b]{Nikhef and the University of Amsterdam, Science Park, 1098XG Amsterdam, Netherlands}

\emailAdd{maximinio.adrover@physik.uzh.ch}
\emailAdd{alexander.bismark@physik.uzh.ch}
\emailAdd{m.flierman@nikhef.nl}

\abstract{The Hamamatsu R12699-406-M4 is a $2\times2$ multi-anode 2-inch photomultiplier tube that offers a compact form factor, low intrinsic radioactivity, and high photocathode coverage. These characteristics make it a promising candidate for next-generation xenon-based direct detection dark matter experiments, such as XLZD and PandaX-xT. We present a detailed  characterization of this photosensor operated in cold xenon environments, focusing on its single photoelectron response, dark count rate, light emission, and afterpulsing behavior. The device demonstrated a gain exceeding \SI{2e6}{} at the nominal voltage of \SI{-1.0}{kV}, along with a low dark count rate of \SI{0.4(2)}{Hz/cm^2}. 
Due to the compact design, afterpulses exhibited short delay times, resulting in some cases in an overlap with the light-induced signal. To evaluate its applicability in a realistic detector environment, two \model units were deployed in a small-scale dual-phase xenon time projection chamber. The segmented $2\times2$ anode structure enabled lateral position reconstruction using a single photomultiplier tube, highlighting the potential of the sensor for effective event localization in future detectors.
}


\begin{document}
\maketitle
\flushbottom
    \section{Introduction} \label{sec:intro}

Low-background dual-phase xenon time projection chambers (TPCs) have demonstrated exceptional sensitivity across a range of rare event searches~\cite{ParticleDataGroup:2024cfk,Baudis:2023pzu}. In addition to the ongoing pursuit of particle dark matter~\cite{aprileFirstDarkMatter2023, aalbersFirstDarkMatter2023, PandaX:2024qfu}, these detectors have achieved several milestones, including the observation of the longest directly measured decay via two-neutrino double electron capture in $^{124}$Xe~\cite{aprileObservationTwoneutrinoDouble2019, LZ:2025hud, PandaX-4T:2024fls}. They have also enabled studies in neutrino physics and searches for neutrinoless double-beta decay~\cite{aalbersNextgenerationLiquidXenon2022}. The remarkable improvements in sensitivity over recent decades have been driven by sustained increases in xenon target mass, stringent background reduction strategies, and the implementation of advanced detector technologies, such as new generations of photomultiplier tubes.

Photomultiplier tubes (PMTs) are reliable sensors for detecting the faint vacuum ultraviolet (VUV) scintillation light at \SI{175}{nm}, produced by particle interactions in the liquid xenon (LXe). The current state-of-the-art xenon-based dark matter detectors, XENONnT~\cite{aprileXENONnTDarkMatter2024a}, LUX-ZEPLIN (LZ)~\cite{ aalbersFirstDarkMatter2023}, and PandaX-4T~\cite{PandaX:2024qfu}, employ the 3-inch-diameter Hamamatsu R11410 photomultiplier tube. While widely used, this device presents several limitations, including its relatively large size, limited spatial granularity when deployed in arrays, and a comparatively high projected background contribution in future-generation experiments~\cite{aprileMaterialRadiopurityControl2022, agostiniSensitivityDARWINObservatory2020f} which will impose even stricter radiopurity requirements. Photosensors must also satisfy demanding specifications for noise, stability and light response~\cite{LUNG201232, LOPEZPAREDES201856, Baudis_2013}. To address these challenges, alternative photosensor models are being actively explored. This article presents an extensive characterization of one promising candidate, the Hamamatsu R12699-406-M4, which is a flat panel type multi-anode PMT~\cite{hamamatsuphotonicsk.k.FlatPanelType2020}. It was specifically developed for low-background applications and optimized for radio-purity, as reported in~\cite{Yun:2024oxp}, where a reduction by an order of magnitude of radioactivity from $^{60}$Co and up to ninefold for $^{238}$U was achieved. 
The same study reports characterization results for 54 units tested in a nitrogen gas environment with temperatures as low as \SI{-105}{\celsius}. While the study demonstrates the robustness of the \model PMT and its performance metrics in face of thermal cycling, we further investigate the model's behavior over prolonged periods of time in its intended application, i.e., submerged in liquid and gaseous xenon at temperatures down to \SI{-100}{\celsius} and pressures of about \SI{2}{\bar}. Here we present our characterization results, as well as a long term stability study and a first application in a small scale detector, exploiting the separate anode read-out capability of the \model PMT.

The device is introduced in \autoref{sec:pmt}. Its characterization, covering the single photoelectron response, dark counts and light emission, afterpulsing, long term stability, and the general quality of the device, was performed in cryogenic xenon environments at the MarmotX facility of the University of Zurich. The measurements at this facility are summarized in \autoref{sec:characterization}. To evaluate the suitability of this PMT model for its intended application, and to assess its position reconstruction capabilities enabled by the segmented anode readout, two units were deployed in the small-scale XAMS TPC located at Nikhef in Amsterdam. Details of this setup and the corresponding studies are provided in \autoref{sec:tpc}. The resulting conclusions and future directions are discussed in \autoref{sec:discussion_conclusion}. The readout circuit used in this work, along with efforts to optimize the radiopurity of this PMT model, are described in \autoref{sec:darwin_2inch_readout} and \autoref{sec:gator}, respectively. Finally, the various origins of the observed dark counts are examined in detail in \autoref{sec:dc_populations}.
    
    \section{The Hamamatsu R12699-406-M4 Photomultiplier Tube} \label{sec:pmt}

The R12699-406-M4 PMT by Hamamatsu Photonics K.K.~\cite{hamamatsuphotonicsk.k.FlatPanelType2020} has a square cross-section with an edge length of \SI{56}{mm}, incorporating a bialkali photocathode with an active area of $\SI{48.5}{mm} \times \SI{48.5}{mm}$. This corresponds to a photocathode coverage of approximately \SI{75}{\percent}, offering a significant improvement over the \SI{62}{\percent} coverage of the circular R11410 PMTs (at their densest array packing), and the \SI{64}{\percent} coverage of the square R8520 model (employed in XENON10 and XENON100~\cite{aprileDesignPerformanceXENON102011, aprileXENON100DarkMatter2012}). The higher photocathode coverage enhances the overall light collection efficiency in TPCs instrumented with this sensor.

\begin{figure}[!t]
	\sbox\twosubbox{%
		\resizebox{\dimexpr.99\textwidth-1em}{!}{%
			\includegraphics[trim={0 0 0 0},clip,height=3cm]{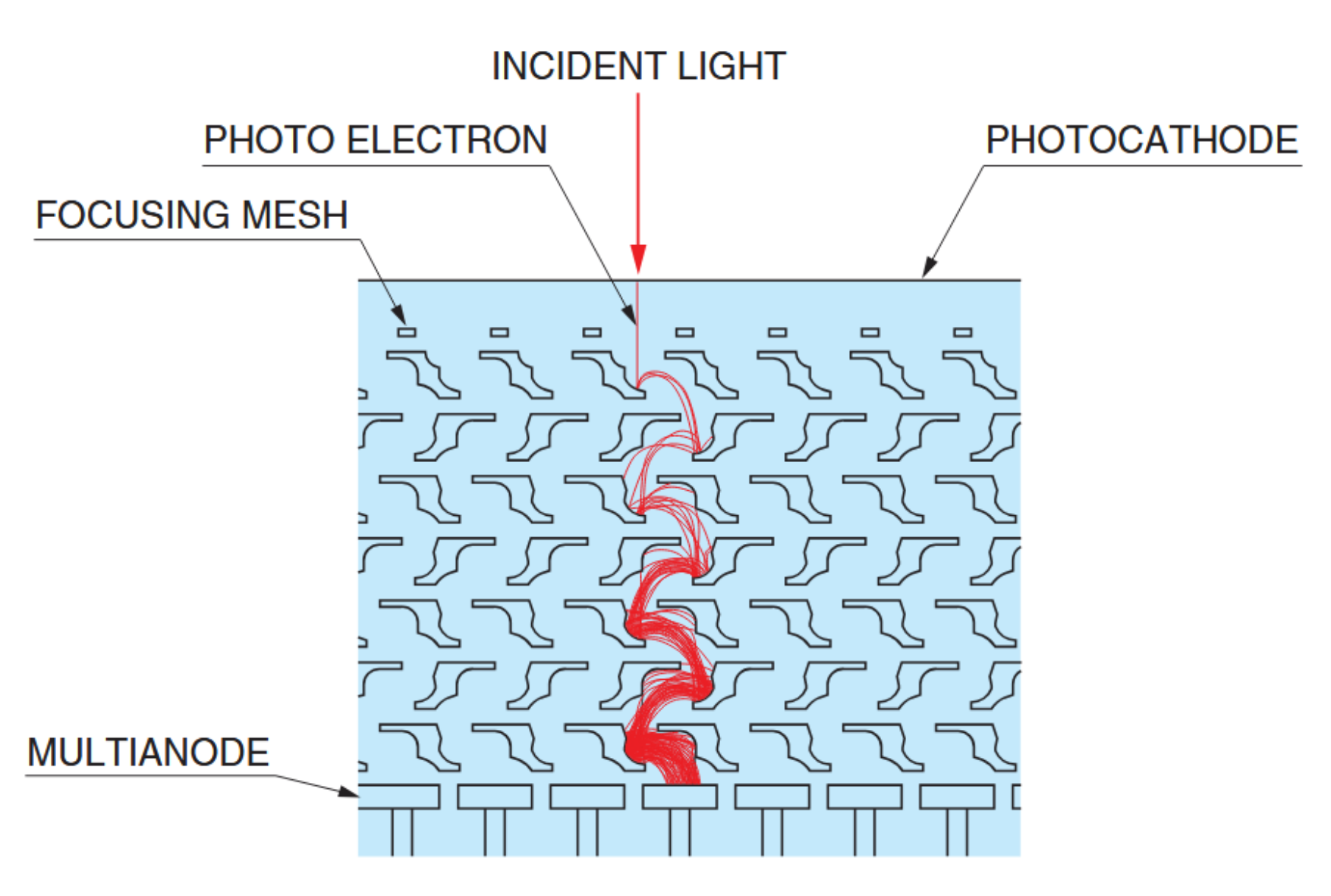}%
			\includegraphics[trim={0 0 0 0},clip,height=3cm]{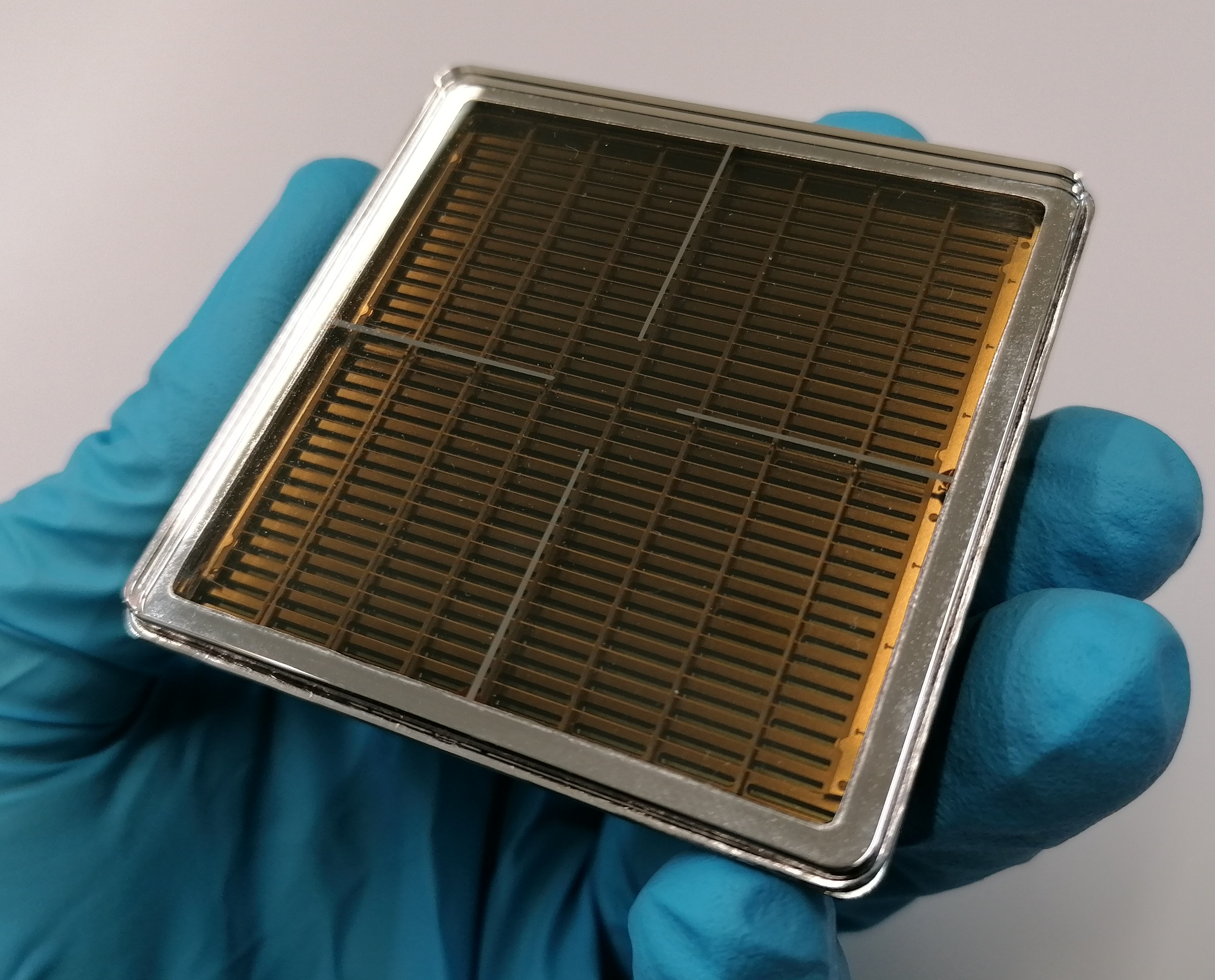}%
		}%
	}
	\setlength{\twosubht}{\ht\twosubbox}
	
	\centering	
	
	\subcaptionbox{}{%
		\includegraphics[trim={0 0 0 0},clip,height=0.95\twosubht]{figures/metal_channel_dynode_pmt}%
	}\qquad
	\subcaptionbox{}{%
		\includegraphics[trim={0 0 0 0},clip,height=0.95\twosubht]{figures/R12699_photo_polished}%
	}
	
	\caption[Hamamatsu R12699-406-M4 PMT structure.]{Hamamatsu R12699-406-M4 PMT structure. \textbf{(a)}~Electrode structure and electron trajectories in a multianode metal channel dynode PMT. Figure provided by~\cite{hamamatsuphotonicsk.k.PhotomultiplierTubesBasics2017}. \textbf{(b)}~Photograph of a Hamamatsu R12699-406-M4 PMT with the first dynode stage visible. The gray crosshair-like feature mitigates charge-up of the photocathode.}\label{fig:r12699}
\end{figure}

The \model PMT employs a metal channel dynode structure for electron multiplication, similarly to the Hamamatsu R8520 PMT. As illustrated in \autoref{fig:r12699}, this configuration consists of densely packed fine dynode channels at each amplification stage, arranged in close proximity. This design enables a compact form factor, resulting in a total PMT  height of \SI{1.5}{cm} (excluding stem pins), making it approximately 7.7 times shorter than the R11410 model. The reduced profile significantly lowers the buoyant force experienced in liquid xenon, by nearly two orders of magnitude compared to the R11410 PMT, thereby allowing for a less rigid mechanical support structure when assembling arrays of this sensor in a LXe TPC. This, in turn, minimizes the amount of material placed near the sensitive xenon volume, which can contribute to a reduction in the overall radioactive background. Additionally, the shorter PMT length reduces the volume of uninstrumented xenon between adjacent photosensors, further enhancing detector efficiency.  

An additional advantage  of the metal-channel dynode structure and the resulting compact design is the fast signal response, characterized by a transit time of \SI{5.9}{ns} and a transit time spread (TTS) of \SI{0.4}{ns}~\cite{hamamatsuphotonicsk.k.FlatPanelType2020}. These fast timing characteristics could be exploited for advanced waveform analyses, including potential pulse shape discrimination techniques to separate nuclear and electronic recoils. The R12699-406-M4 model has $n=10$ dynode stages and a multianode configuration consisting of a $2\times2$ matrix, with each sector measuring $\SI{24.25}{mm} \times \SI{24.25}{mm}$. The anodes can be read out individually or in combination, offering flexibility in optimizing channel count and spatial granularity within an array. This enables, for example, the use of finer granularity near the TPC walls for improved edge resolution, while coarser granularity can be applied in central regions to reduce readout complexity. Details of the voltage divider circuit (referred to as the \textit{base} in the following) used in this work, for both combined and separate anode readout modes, are provided in \autoref{sec:darwin_2inch_readout}.

These properties, combined with the high sensitivity of the PMT to xenon scintillation light, featuring a quantum efficiency (QE) of \SI{33}{\percent}~\cite{hamamatsuphotonicsk.k.FlatPanelType2020} at VUV wavelengths equivalent to that of the R11410-21 PMTs~\cite{antochiImprovedQualityTests2021}, its low intrinsic radioactivity (see \autoref{sec:gator}), and its rated operation down to \SI{-110}{\celsius}, make the R12699-406-M4 a strong candidate for future xenon-based dark matter detectors. Accordingly, it has recently been selected as the primary photosensor option for the future PandaX-xT experiment~\cite{abdukerimPandaXxTMultitentonneLiquid2024}.

    \section{Characterization of the Photomultiplier} \label{sec:characterization}

To evaluate the suitability of the Hamamatsu R12699-406-M4 PMT for use in LXe TPCs, its performance must be assessed under conditions that closely replicate those within such detectors. This includes not only reference measurements in vacuum at room temperature, but also tests conducted in xenon gas and liquid at cryogenic temperatures around \SI{-100}{\celsius} and pressures of approximately \SI{2}{bar}. These operating conditions are achievable at the MarmotX photosensor testing facility at the University of Zurich~\cite{antochiImprovedQualityTests2021, adroverCharacterizationNovelHamamatsu2023, bismarkTestsFundamentalsQuantum2024}, where the primary characterizations reported in this work were carried out. The study focused on key PMT properties relevant to rare event searches, including the single photoelectron response, dark count rate, spontaneous light emission, afterpulsing behavior, and long term stability.

\subsection{The MarmotX Facility: Setup and Measurement Campaigns}
\label{sec:marmotx}

The MarmotX facility comprises a double-walled cryostat connected to an external gas handling system. The cooling required to liquefy xenon is provided by a pulse tube refrigerator. The operating temperature is monitored constantly and controlled by means of a Cryocon 32b cryogenic temperature controller. Over the duration of the runs the temperature of the setup consistently remained stable within a few \unit{\milli\kelvin}, ensuring that systematic changes in PMT response due to ambient effects are negligible. In the primary characterization campaigns of the Hamamatsu R12699-406-M4 PMT conducted at MarmotX, up to four units were simultaneously operated within the cryostat. For the most recent long-term stability measurements, the setup  was upgraded to accommodate up to eight photosensors. 

\begin{figure}[!t]
	\sbox\twosubbox{%
		\resizebox{\dimexpr.99\textwidth-1em}{!}{%
			\includegraphics[trim={0 0 0 0},clip,height=3cm]{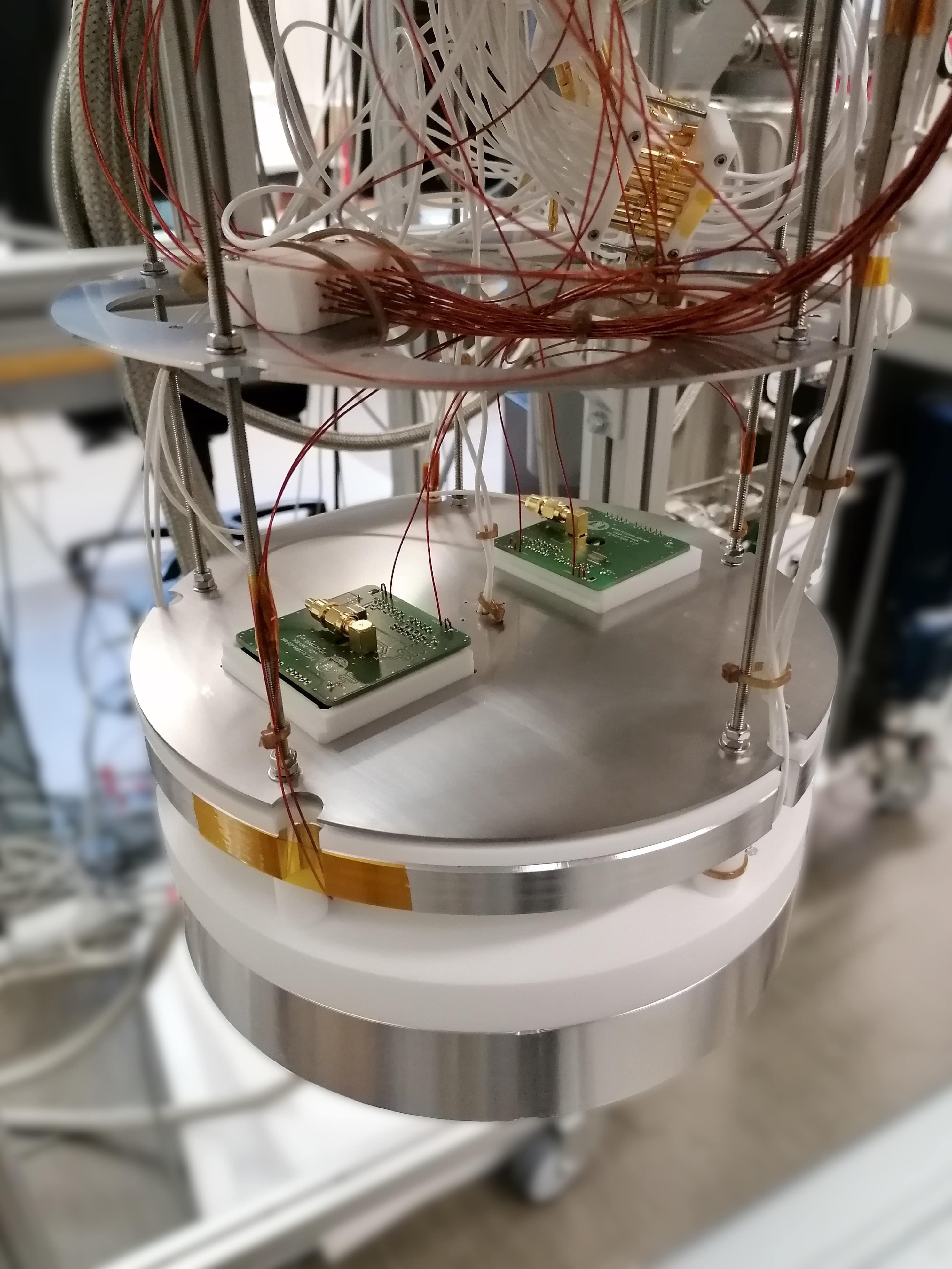}%
			\includegraphics[trim={0 0 0 0},clip,height=3cm]{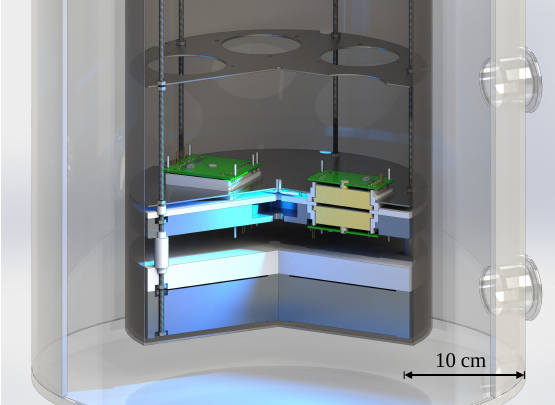}%
		}%
	}
	\setlength{\twosubht}{\ht\twosubbox}
	\centering	
		\subcaptionbox{}{%
		\includegraphics[trim={0 0 0 0},clip,height=0.90\twosubht]{figures/Marmotx_setup_edited-compressed}%
	}\quad
	\subcaptionbox{}{%
		\includegraphics[trim={0 0 0 0},clip,height=0.90\twosubht]{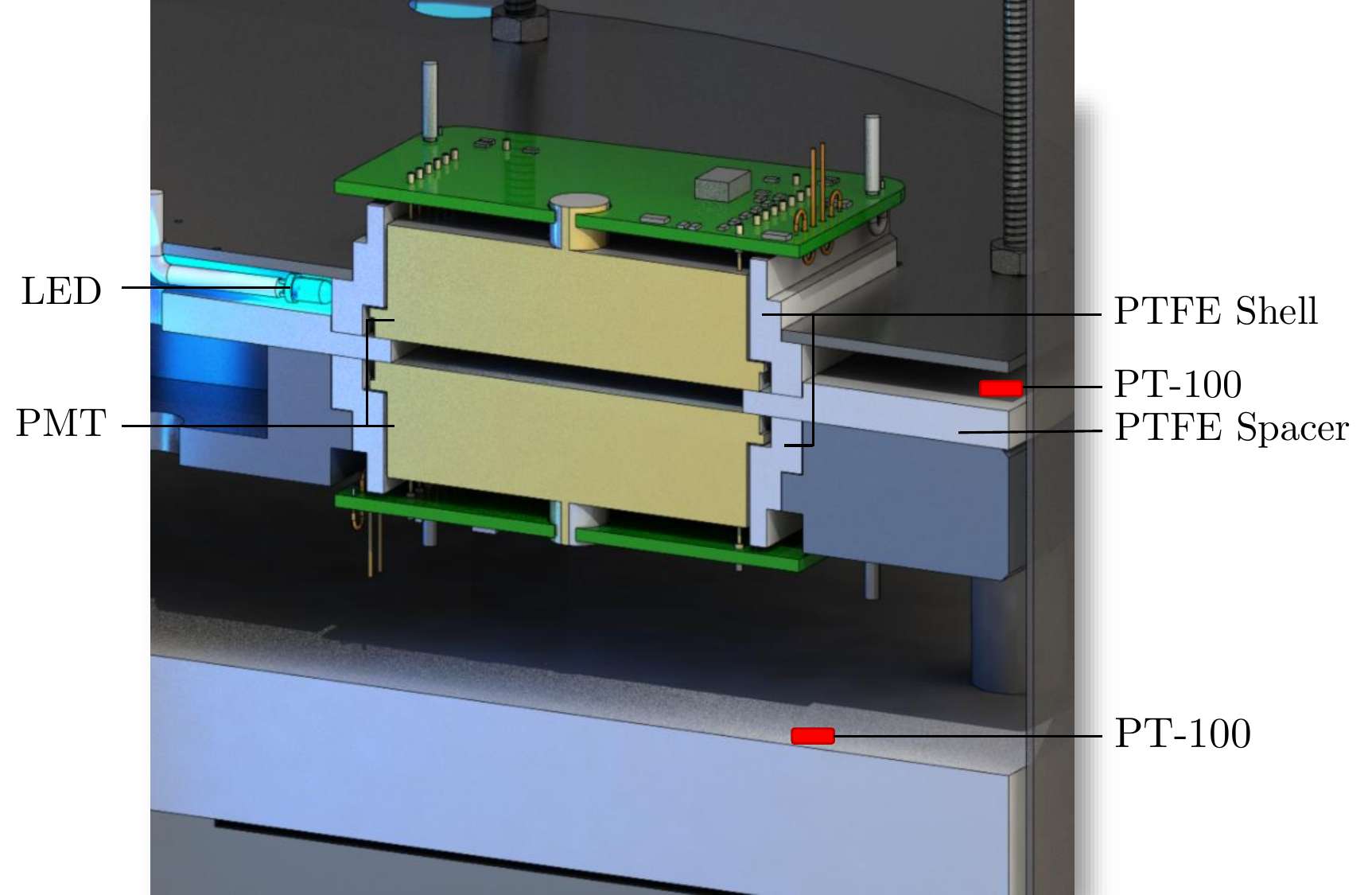}%
	}
	\caption[Holder setup for R12699-406-M4 PMT characterizations in MarmotX.]{Holder setup for the simultaneous characterization of four Hamamatsu R12699-406-M4 PMTs in MarmotX. Two PMTs each are paired with facing windows at a \SI{3}{mm} distance. \textbf{(a)}~Photograph of the assembled setup during preparation for the third data-taking campaign. \textbf{(b)}~CAD cutaway rendering of the assembly inside the double-walled cryostat. The blue light is illustrated by the blue hue surrounding the LED.}\label{fig:marmotx_setup}
\end{figure} 

The PMT support structure, shown in \autoref{fig:marmotx_setup}, consists of stacks of discs made of PTFE, aluminum, and stainless steel, with a temperature sensor placed below the bulk, and another one towards the edge of the setup at the same height as the LED. The PMTs are mounted in closely spaced pairs, with their windows facing each other at a distance of \SI{3}{mm}. This arrangement enables mutual monitoring for potential light emission from the opposite sensor (see \autoref{sec:characterization_dc_le}) and allows investigation of temporal signal correlations, aiding in the classification of dark count populations (see \autoref{sec:dc_populations}). All PMTs were operated at their nominal voltage of \SI{-1.0}{kV} and, in some temporary cases, at up to \SI{-1.1}{kV}, as advised and permitted by the manufacturer specifications~\cite{hamamatsuphotonicsk.k.FlatPanelType2020}. The HV is supplied to the PMTs by two four-channel CAEN NV1470 power supply units. Each PMT is enclosed in a PTFE shell that electrically insulates the casing, which sits at the cathode potential. A thin PTFE disk ensures accurate spacing between paired sensors. 
For optical stimulation, each PMT pair can be illuminated independently by a blue LED ($\lambda_p =\SI{468}{\nano\meter})$ placed at a short distance from the gap between PMT windows, as depicted in~\autoref{fig:marmotx_setup}. The LED-light reaching the PMTs is diffused by the enveloping PTFE shells, such as to avoid spot-like illumination of the photocathode. In the last data-taking campaign (Run 3), a single LED source was coupled to all units via an optical fiber bundle. The LED is pulsed by a waveform generator, which provides a synchronous trigger signal to the data acquisition system. The pulse shape is characterized by a \SI{5}{\nano \second} rise and fall and \SIrange{30}{50}{\nano \second} (\SI{100}{\nano\second}) in the single LED (fiber) configuration. 
Given the fast response  of the PMT model, signals were digitized using a 14-bit CAEN V1730D ADC at a sampling rate of \SI{500}{MS/s}, following tenfold amplification by means of a custom-built~\cite{JulienWulf} dual-gain (0.5- and 10-fold) amplification module with a 250 MHz bandwidth in the high-gain stage. 
\begin{table}[t]
    \centering
    \caption[Overview of MarmotX data acquisition campaigns.]
    {Overview of the MarmotX data acquisition campaigns. The duration of each run is given in days, with operation in gaseous xenon (GXe) and liquid xenon (LXe) indicated separately. The PMTs installed during each campaign are listed by manufacturer ID, grouped by facing pairs (columns). PMTs that were installed but not operational for the full run are indicated in italic. Units with the “MB” prefix denote an improved low-radioactivity version, as discussed in \autoref{sec:gator}.}
    \label{tab:runs}
    \begin{tabular}{@{}c c cccc@{}}
        \toprule
        \textbf{Run} & \textbf{Duration [d]} & \multicolumn{4}{c}{\textbf{PMT Pairs (Facing Windows)}} \\
        & \textbf{(Total / GXe / LXe)} & Pair 1 & Pair 2 & Pair 3 & Pair 4 \\
        \midrule
        \multirow{2}{*}{0} & \multirow{2}{*}{\phantom{0}71 / -- / 14} 
            & MA0055 &  &  &  \\
         &  & MA0058 &  &  &  \\
        \midrule
        \multirow{2}{*}{1} & \multirow{2}{*}{\phantom{0}89 / 9 / 13} 
            & MA0055 & MB0072 &  &  \\
         &  & MA0058 & \textit{MB0080} &  &  \\
        \midrule
        \multirow{2}{*}{2} & \multirow{2}{*}{100 / 7 / 41} 
            & MB0090 & \textit{MB0113} &  &  \\
         &  & MB0015 & \textit{MB0112} &  &  \\
        \midrule
        \multirow{2}{*}{3} & \multirow{2}{*}{\phantom{0}87 / -- / 62} 
            & \textit{MB0080} & MA0058 & MB0015 & MB0113 \\
         &  & MA0055 & \textit{MB0072} & MB0112 & \textit{MB0090} \\
        \bottomrule
    \end{tabular}
\end{table}
The R12699-406-M4 PMT characterizations program at the MarmotX facility was conducted over four distinct data-taking campaigns, referred to as Run~0 -- Run~3. 
The duration of each run and the specific PMT units involved are summarized in \autoref{tab:runs}. In total, eight R12699-406-M4 PMTs were installed in the facility at least once. 
Run~0 involved only two PMTs and is excluded from detailed discussion here due to unstable facility conditions. In Runs~1 and 2, four PMTs were installed simultaneously for baseline performance characterization. During Run~3, all eight units were mounted for long-term stability testing.
Not all PMTs remained operational throughout their respective campaigns due to connectivity issues or damage to the voltage divider circuits. Units excluded from analysis are marked accordingly in \autoref{tab:runs}.
Notably, PMT MB0080 exhibited a progressive increase in output pulse width over time, leading to a reduced pulse amplitude that eventually became indistinguishable from noise. This behavior was traced to a dramatic increase in the output impedance and, thus, time constant of its base and was observed consistently over all runs. A similar effect was observed for PMT MB0113 during Run~2, though it did not recur in Run~3. 

The room-temperature reference measurements presented below were performed in vacuum at pressures on the order of $\mathcal{O}(10^{-7}-10^{-5})$\;mbar simulating the conditions inside a TPC prior to xenon gas filling. Measurements in GXe and LXe were conducted at pressures of approximately \SI{2}{bar} and temperatures near \SI{-100}{\celsius}, varying slightly from run to run, matching typical operating conditions of dual-phase xenon detectors.

\subsection{Single Photoelectron Response} 
\label{sec:characterization_spe}

One of the key performance characteristics investigated for the Hamamatsu R12699-406-M4 PMT is its response to low light levels at the single photoelectron (SPE) level. Assuming sufficient gain linearity, these results can be extrapolated to higher illumination levels. The use of a pulsed LED, synchronized with an external trigger, enables a fixed signal integration window for precise charge measurements. The PMT gain is extracted using the model-independent statistical approach described in~\cite{saldanhaModelIndependentApproach2017}. This method foregoes any underlying assumption on specific shapes of the PMT charge spectrum and naturally includes contributions from underamplified photoelectrons. The minimal relative uncertainty of this method is achieved at an occupancy $\lambda $, denoting the expected number of photoelectrons produced per trigger, of $\approx \SI{2}{PE/trigger}$, set by appropriate tuning of the LED pulse amplitude and width.
\begin{figure}[t!]
	\sbox\twosubbox{%
		\resizebox{\dimexpr.99\textwidth-1em}{!}{%
			\includegraphics[trim={7 0 0 0},clip,height=3cm]{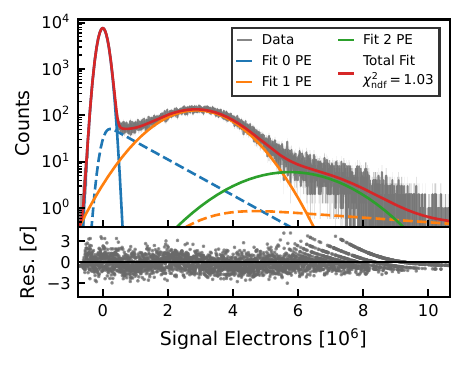}%
			\includegraphics[trim={0 0 7 -2},clip,height=3cm]{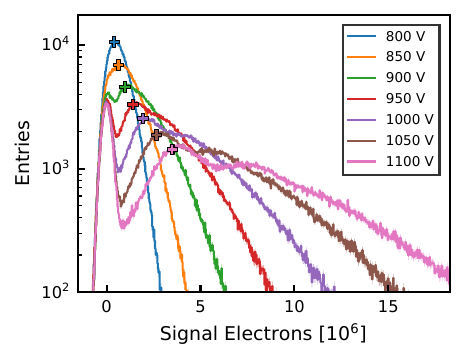}%
		}%
	}
	\setlength{\twosubht}{\ht\twosubbox}
	
	\centering	
	
	\subcaptionbox{}{%
		\includegraphics[trim={7 0 0 0},clip,height=0.99\twosubht]{figures/spe_spectrum_fit_example}%
	}\quad
	\subcaptionbox{}{%
		\includegraphics[trim={0 0 7 -2},clip,height=0.99\twosubht]{figures/roomtemp_hvscan_areaspectra_MA0055}%
	}	
	\caption[R12699-406-M4 PMT charge spectra.]{\textbf{(a)}~Example of an analytical model fit~\cite{diwanStatisticsChargeSpectrum2020} to a charge spectrum of PMT MA0058 at room temperature and nominal bias voltage of \SI{-1000}{\kilo\volt}. Gaussian components are shown as solid lines, exponentially modified Gaussian contributions as dashed lines. The fit yields a gain of $\mu\approx 2.9\cdot10^6$, occupancy of $\lambda\approx 0.13$\;PE/trigger, relative exponentially modified Gaussian contribution $w \approx 3\;\%$, and SPE resolution of $\sigma/\mu\approx 36\;\%$. The fit also gives  a peak-to-valley ratio of approximately 2.5. The residuals to the fit, in units of one standard deviation, are shown in the bottom plot. \textbf{(b)}~Room-temperature charge spectra of PMT MA0055, for various bias voltages under constant illumination. Markers indicate gain estimates from the model-independent approach.}\label{fig:r12699_spectra}
\end{figure}

The ability of the PMT to separate signal from background is typically characterized by the \textit{SPE-resolution} and the \textit{peak-to-valley ratio}. The latter is not accessible via the model-independent approach. Therefore, additional charge spectra were acquired at a lower occupancy $\lambda$ of $\mathcal{O}(0.1)$\;PE/trigger and were fitted with an analytical model developed in~\cite{diwanStatisticsChargeSpectrum2020}, to report on the SPE-resolution and peak-to-valley ratio. As these quantities are dependent on the gain, and hence the bias voltage, the reported results were computed from measurements at bias voltages resulting in gain values of \num{\ca 2e6} that may vary slightly from PMT to PMT.
An exemplary fit of this model to a R12699-406-M4 PMT charge spectrum recorded at room temperature is shown in \autoref{fig:r12699_spectra}~(a). It demonstrates its composition of a series of equidistant Gaussian peaks with Poissonian weighted amplitudes according to the occupancy, corrected by supplementary exponentially modified Gaussian contributions.

The SPE signal resolution, defined as the ratio of the standard deviation to the mean of the SPE response, extracted from the fit, was found to be approximately independent of high voltage at gains of $\gtrsim\SI{2e6}{}$. We measured a mean SPE resolution of \SI{37(2)}{\percent} at room temperature and \SI{35(3)}{\percent} in LXe. 
This is approximately \SI{1.23(14)}{} higher than reported for R11410-21 PMTs at the same gain~\cite{barrowQualificationTestsR11410212017}. 

The peak-to-valley ratio, defined as the ratio of the SPE peak maximum height and the valley between this peak and the noise pedestal, was observed to increase with applied bias voltage, saturating at gains around \SI{4e6}{}. At an occupancy of $\ca \SI{0.04}{PE/trigger}$, the R12699-406-M4 PMTs exhibit a peak-to-valley ratio of \SI{2.2(4)}{}.
While the analytical fit also allows gain extraction via the mean of the SPE Gaussian, it neglects underamplified events and overestimates the gain by roughly \SI{10}{\percent}. Both methods were used for cross-checks, but only gain values from the model-independent approach are used hereafter for consistency. 

\begin{figure}[!t]
	\sbox\twosubbox{%
		\resizebox{\dimexpr.99\textwidth-1em}{!}{%
			\includegraphics[trim={7 0 0 0},clip,height=3cm]{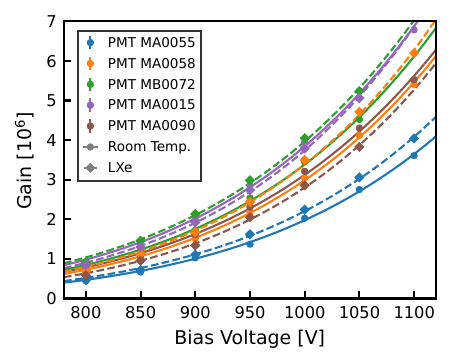}%
			\includegraphics[trim={0 0 7 -2.75},clip,height=3cm]{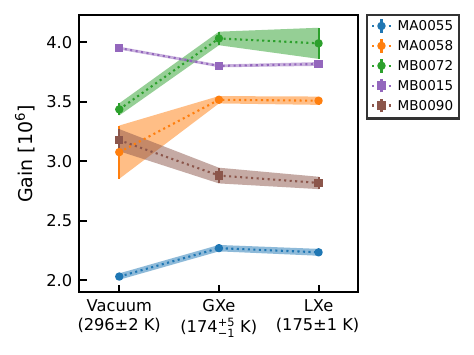}%
            \label{subfig:r12699_hv_dep_fit}
		}%
	}
	\setlength{\twosubht}{\ht\twosubbox}
	
	\centering	
	
	\subcaptionbox{}{%
		\includegraphics[trim={7 0 0 0},clip,height=0.99\twosubht]{figures/comp_hvscan_hvgainfits_runs}%
	}\quad
	\subcaptionbox{}{%
		\includegraphics[trim={0 0 7 -2.75},clip,height=0.99\twosubht]{figures/gains_comparison_conditions}%
	}
	
	\caption[R12699-406-M4 PMT gain dependence on HV and environment.]{R12699-406-M4 PMT gain dependence on the bias voltage and environment. \textbf{(a)}~Gain as a function of high voltage fitted with a power law \autoref{eq:gain_hv_power_law}. \textbf{(b)}~Gain measurements across different environments. In Run~2 (squares), expected gain increases upon cool-down were not observed due to systematic bias introduced by LED pulser switching noise.}\label{fig:r12699_gain_conditions}
\end{figure}

The gain dependence on the magnitude of high voltage $U$, shown in \autoref{fig:r12699_spectra} and \autoref{fig:r12699_gain_conditions}, follows the power law
\begin{align}
	G(U) = \left(a \cdot \left(\frac{E}{\SI{1}{\volt}}\right)^k \right)^n = a^n \left(\frac{U}{\SI{1}{\volt}} \cdot \dfrac{1}{n+1}\right)^{kn} \equiv A \cdot U^{kn}, \label{eq:gain_hv_power_law}
\end{align} 
where $a$ is an empirical parameter, $E$ is the interstage voltage of dynodes, $n=10$ is the number of dynode stages, and $k$ is a PMT-specific constant determined by the dynode structure and material.
Fit parameters yielded average values of $a=\SI{0.255(6)}{}$ and $k=\SI{0.632(3)}{}$ ($a=\SI{0.254(13)}{}$) and $k=\SI{0.637(9)}{}$) at room temperature (in LXe). All PMTs exceeded the manufacturer-specified gain of \SI{1.5e6}{} at \SI{-1.0}{kV}, with an average measured gain of \SI{3.3e6}{} and a standard deviation of \SI{0.7e6}{}.

In Run~1, a gain increase of \SI{14(2)}{\percent} was observed upon cooling in GXe, consistent with behavior reported for R11410-21 PMTs~\cite{barrowQualificationTestsR11410212017}. Similar increases were observed in Run~0. However, in Run~2, a gain decrease was observed in the two operational units. This was attributed to baseline distortions caused by increased required LED drive voltage and resulting switching noise, possibly following LED displacement from thermal contraction. As a result, gain estimates in Run~2 reflect relative stability within a single environment, but not absolute comparisons across conditions.

\begin{figure}[!t]
	\sbox\twosubbox{%
		\resizebox{\dimexpr.99\textwidth-1em}{!}{%
			\includegraphics[trim={7 0 0 0},clip,height=3cm]{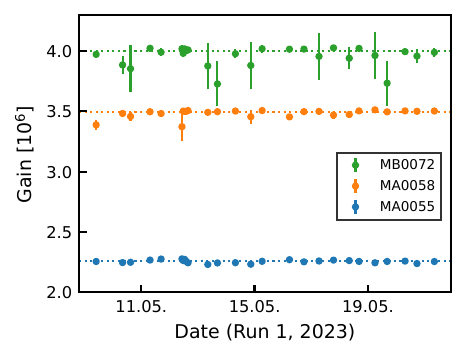}%
			\includegraphics[trim={0 0 7 0},clip,height=3cm]{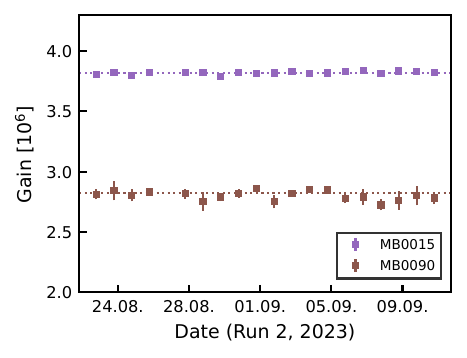}%
		}%
	}
	\setlength{\twosubht}{\ht\twosubbox}
	
	\centering	
	
	\subcaptionbox{}{%
		\includegraphics[trim={7 0 0 0},clip,height=0.99\twosubht]{figures/gains_run1_stability_lxe}%
	}\quad
	\subcaptionbox{}{%
		\includegraphics[trim={0 0 7 0},clip,height=0.99\twosubht]{figures/gains_run2_stability_lxe}%
	}

    \subcaptionbox{}{%
		\includegraphics[trim={0 0 0 0},clip,height=0.99\twosubht]{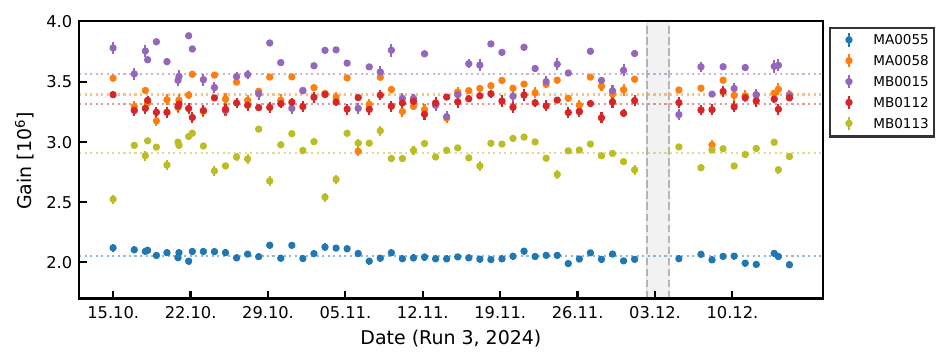}%
	}
	
	\caption[R12699-406-M4 PMT gain stability in liquid xenon.]{Gain evolution during long-term stability tests in LXe (a) Run~1, (b) Run~2, and (c) Run~3. Median gain values for each PMT are indicated by a dotted line. Calibrations recorded within the shaded area in (c) had to be omitted due to a malfunction of the DAQ.}\label{fig:r12699_gain_evolution_lxe}
\end{figure}

For rare event searches, relying on an optimal resolution over extended data taking periods, stable performance of the sensors is a prerequisite. Long term drifts of the SPE response are a negative feature in this kind of experiment. To gauge the long term behavior of the \model PMT, gain stability campaigns in LXe were conducted toward the end of Runs~1 and 2, spanning \SI{13}{days} and \SI{20}{days}, respectively. During these periods, the PMTs were operated continuously at constant bias voltage, with gain calibrations performed at least once per day. 

As shown in \autoref{fig:r12699_gain_evolution_lxe}~(a,b), the gain evolution was stable within the measurement uncertainties, with a typical per-PMT gain variation of \SI{1.0(6)}{\percent}. In Run~3, a long-term stability study was carried out with five operational PMTs over a period of \SI{62}{days}, as illustrated in \autoref{fig:r12699_gain_evolution_lxe}~(c). Compared to gain values obtained in the previous runs, a decrease in the average gain of all units of \SI{-5.9(1.9)}{\percent} and an increase in gain fluctuations were observed. The gain of each unit remained stable within \SI{4.9}{\percent}. Gain variations beyond \SI{3}{\percent} were consistently observed in the units located in the top position of their respective PMT pair. These were connected to a separate power supply unit (PSU) compared to the units at the bottom of the pair. The two units connected to a second PSU both exhibited fluctuations below \SI{1.8}{\percent}. As the monitored voltage and current of the PSUs were not recorded, systematic effects arising from baseline distortions cannot be excluded. Furthermore, at low light intensities the gain is expected to be independent of the number of incident photons, and hence, the occupancy. A feature of the chosen estimation method, however, is an explicit anti-correlation between the SPE response and the estimated occupancy~\cite{saldanhaModelIndependentApproach2017}. The observed correlation coefficient of the relative gain and occupancy changes between subsequent calibrations of \num{-.9}, therefore indicates that the observed short-term fluctuations arise from fluctuations in the occupancy estimate, the precision of which is greatly affected by unstable noise conditions.

To evaluate the long-term stability of the individual gains, a linear fit is applied to the calibration data of each PMT. As the uncertainty on the fit is dominated by the relatively large short-term fluctuations of the estimates, a jackknife resampling approach was employed. The mean and variance of the relative changes between the first and last point of the respective fits are calculated over all possible subsamples with two data points removed. For all units, the estimated means of the relative changes agree with zero within one standard deviation. While the random fluctuations on individual gain estimates have increased, there is no indication of long-term effects affecting the gain of all tested units under extended cryogenic operation.

\subsection{Dark Count Rates and Light Emission} \label{sec:characterization_dc_le}

Rare event searches require minimal background noise, particularly in low-energy regions, to ensure sensitivity to faint signals. A key limiting factor is the rate of \textit{accidental coincidences} (ACs), i.e., random, uncorrelated signals detected simultaneously by multiple photosensors. The primary contributor to such ACs are \textit{dark counts} (DCs), which refer to spontaneous PMT pulses occurring in the absence of incident light. These dark counts typically originate from thermionic electron emission from the photocathode and, to a lesser extent, from dynode surfaces~\cite{hamamatsuphotonicsk.k.PhotomultiplierTubesBasics2017}. Their rate (DCR) increases with both supply voltage and temperature and can be substantially reduced by cryogenic cooling, as is naturally achieved in LXe TPCs. However, at high bias voltages, additional emission from dynodes in strong electric fields may further elevate the DCR~\cite{hamamatsuphotonicsk.k.PhotomultiplierTubesBasics2017}. 

Moreover, DCR estimates can also be influenced by scintillation or Cherenkov radiation generated in the PMT glass due to internal radioactivity (e.g., $^{40}$K decays), gamma emission from surrounding materials, or cosmic rays. As such, measured rates in the MarmotX facility may exceed those expected in shielded, low-background experimental environments. A detailed breakdown of the dark count components and their respective origins is provided in \autoref{sec:dc_populations}.

An additional complication arises from  \textit{light emission}, a phenomenon in which single photons are emitted from internal PMT structures, typically at elevated voltages~\cite{AKIMOV20151}. This light can be detected by the emitting PMT itself or by neighboring units, especially in the face-to-face configuration used in MarmotX~\cite{antochiImprovedQualityTests2021}. The most common form, known as \textit{micro light emission}, manifests itself as a low-intensity photon flux that increases with supply voltage. It can be identified as a bias-dependent offset in DCR and diagnosed by monitoring the count rate while varying the HV on the opposing PMT. 

Two triggering modes were employed in MarmotX to study dark counts and light emission.
For continuous time-dependent DCR measurements, the rate of trigger signals from a discriminator, using a threshold corresponding to approximately $\nicefrac{1}{4}$\;PE, was recorded in 1-second intervals. This approach is particularly useful to investigate HV-dependent light emission.
However, this method alone is insufficient for precise absolute DCR measurements. During initial estimations performed before the main characterization campaigns, it was observed that discriminator-triggered and ADC self-triggered acquisitions underestimated the DCR due to fixed thresholds and, hence, susceptibility to baseline shifts. To overcome these limitations, absolute rate estimates are based on waveform data acquired randomly, without requiring signal presence. The fraction of waveforms exceeding a fixed threshold per PMT (typically $\nicefrac{1}{4}$~PE) is Poisson-corrected for the probability of multiple dark counts per waveform and converted to a DCR based on the total acquisition time~\cite{bismarkPMTAnalysis2023}. 

\begin{figure}[!t]
	\sbox\twosubbox{%
		\resizebox{\dimexpr.99\textwidth-1em}{!}{%
			\includegraphics[trim={7 0 0 0},clip,height=3cm]{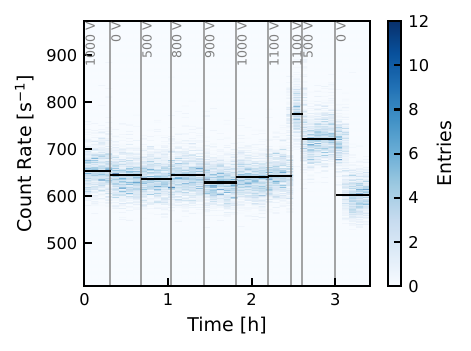}%
			\includegraphics[trim={0 -14 7 -3.5},clip,height=3cm]{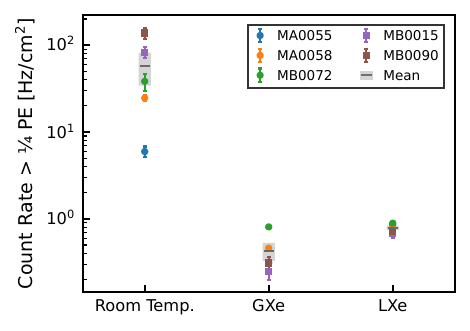}%
		}%
	}
	\setlength{\twosubht}{\ht\twosubbox}
	
	\centering	
	
	\subcaptionbox{}{%
		\includegraphics[trim={7 0 0 0},clip,height=0.99\twosubht]{figures/scaler_channel_3_rate_vs_t_h_rel_hist2d_1678972178-1678984437_mode}%
	}\quad
	\subcaptionbox{}{%
		\includegraphics[trim={0 -14 7 -3.5},clip,height=0.99\twosubht]{figures/DC_rates_runs_simple}%
	}
	
	\caption[Light emission and dark count rates R12699-406-M4 PMT.]{Count rates of R12699-406-M4 PMTs. \textbf{(a)}~Coun rate evolution in PMT MB0072 at room temperature, showing increased rates when the opposing PMT MB0080 is biased at \SI{-1.1}{kV}. The observed \SI{21}{\percent} mode rate increase is attributed to light emission from MB0080. \textbf{(b)}~Measured rates above $\nicefrac{1}{4}$~PE in different environments. Cooling in GXe significantly suppresses thermionic emission, while the elevated rates in LXe are due to xenon scintillation light.}\label{fig:r12699_dc_rate} 
\end{figure}

Most PMTs showed only statistically insignificant changes in DCR when the HV of the opposing PMT was varied. A notable exception was MB0080, which exhibited signs of light emission. As shown in \autoref{fig:r12699_dc_rate}, MB0072 registered a \SI{21}{\percent} DCR increase approximately \SI{2500}{s} after MB0080 was ramped to \SI{-1.1}{kV}, and a residual \SI{13}{\percent} increase persisted even after reducing the HV of MB0080 to \SI{-0.5}{kV}. After cool-down in GXe, the DCR in MB0072 increased by up to a factor of \SI{12.3(6)}{}, or \SI{216(6)}{Hz}, when MB0080 was powered versus disabled. Light emission can be an early warning sign of a compromised vacuum. After the conclusion of the campaign, MB0080 was returned to the manufacturer, and later on confirmed to have suffered a leak. 
Mean DCR estimates at nominal voltage for each PMT in room temperature, GXe and LXe are summarized in \autoref{fig:r12699_dc_rate}~(b). As expected, room-temperature values were highest, averaging \SI{60(20)}{Hz/cm^2}. In GXe, thermionic suppression led to a drastic reduction, with a mean of \SI{0.43}{Hz/cm^2} and a standard deviation of \SI{0.20}{Hz/cm^2}. For comparison, R11410-21 PMTs under similar cryogenic conditions exhibited DCRs above $\nicefrac{1}{4}$\;PE of \SI{1.4}{Hz/cm^2} mean and \SI{0.7}{Hz/cm^2} spread~\cite{barrowQualificationTestsR11410212017}, making the R12699-406-M4 values lower by a factor of \SI{3(2)}{}. In LXe, the observed rate approximately doubled compared to GXe, primarily due to xenon scintillation light emitted between the closely spaced PMT windows.

\subsection{Afterpulsing} \label{sec:characterization_ap}

\begin{figure}[!t]
	\sbox\twosubbox{%
		\resizebox{\dimexpr.99\textwidth-1em}{!}{%
			\includegraphics[trim={7 0 0 0},clip,height=3cm]{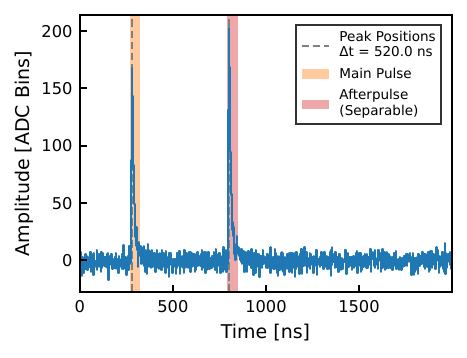}%
			\includegraphics[trim={0 0 7 0},clip,height=3cm]{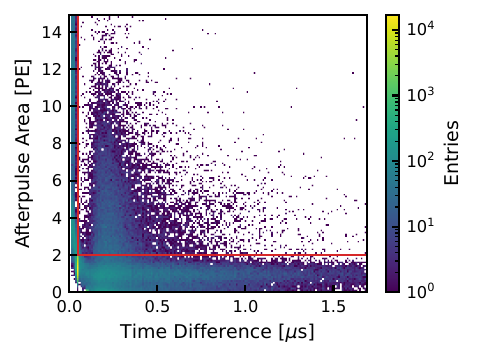}%
		}%
	}
	\setlength{\twosubht}{\ht\twosubbox}
	
	\centering	
	
	\subcaptionbox{}{%
		\includegraphics[trim={7 0 0 0},clip,height=0.99\twosubht]{figures/ap_candidate_wf_example}%
	}\quad
	\subcaptionbox{}{%
		\includegraphics[trim={0 0 7 0},clip,height=0.99\twosubht]{figures/ap_area_vs_tdiff_gxe}%
	}
	
	\caption[Afterpulse studies R12699-406-M4 PMT.]{Afterpulse studies of R12699-406-M4 PMTs in MarmotX. \textbf{(a)}~Exemplary baseline-subtracted and inverted waveform with an identified LED-induced single-PE main pulse and a separable afterpulse candidate, recorded in a cross-check measurement at a low occupancy of 0.1\;PE/trigger for PMT MB0072 at room temperature. \textbf{(b)}~Delay time to the main pulse and area of all reconstructed afterpulse candidates at 8\;PE/trigger illumination of PMT MB0072 in cryogenic GXe.}\label{fig:r12699_afterpulses}
\end{figure}

Afterpulses (APs) are secondary pulses that follow a primary PMT signal, as illustrated in \autoref{fig:r12699_afterpulses}~(a). 
Fast APs originate from electrons elastically backscattering off the first dynode, subsequently returning and triggering another signal on the same dynode ~\cite{hamamatsuphotonicsk.k.PhotomultiplierTubesBasics2017, barrowQualificationTestsR11410212017}. In the R12699 PMTs, these fast APs are typically indistinguishable from the main pulse due to the short time separations. Additional fast structures may arise from temporally extended light emission from the LED itself, motivating the use of minimal pulse widths. Delayed afterpulses are typically caused by ion feedback: residual gas molecules in the PMT are ionized by energetic electrons and subsequently drift toward the photocathode, where they release secondary electrons upon impact. The timing of these APs depends on the electric field geometry and the mass-to-charge ratio of the ions~\cite{mayaniparasPhotomultiplierTubesXENON1T2017}. 
\autoref{fig:r12699_afterpulses}~(b) presents the delay and charge distribution of AP candidates recorded for PMT MB0072 in cryogenic GXe. Due to the confidential internal structure of the R12699-406-M4 PMT, exact AP timing predictions are challenging. However, estimates suggest significantly shorter delays compared to R11410 PMTs. For instance, He$^+$ APs are expected at tens of ns versus \SI{0.52}{\micro\second} in R11410~\cite{barrowQualificationTestsR11410212017}. This property, resulting from the compact geometry, increases the likelihood of APs overlapping with the main pulse, particularly for fast ions and larger primary signals, potentially biasing reconstructed event energy and position.

To characterize AP behavior, MarmotX data were acquired using an external trigger with LED illumination at \SI{8}{PE/trigger}. A peak-finding algorithm~\cite{bismarkPMTAnalysis2023} identifies LED-induced pulses and any subsequent AP candidates. To ensure reliable identification, an AP is labeled as separable if the valley between adjacent peaks drops below \SI{5}{\percent} of the peak height, reducing the likelihood of area overestimation due to pulse overlap. This separability condition effectively isolates true APs from pile-up effects, as seen in \autoref{fig:r12699_afterpulses}(b), where elevated fast AP activity is evident at time differences below \SI{50}{ns}. A distinct population of single-PE events with delays that can surpass \SI{1}{\micro\second} likely corresponds to dark pulses from spontaneous electron emission~\cite{antochiImprovedQualityTests2021}. Another broad population appears around \SI{0.2}{\micro\second}, extending to \SI{\sim10}{PE}, with a decreasing signal size at longer delays, consistent with ion-induced APs from heavier species~\cite{mayaniparasPhotomultiplierTubesXENON1T2017}. This structure is also observed prior to xenon exposure.

During operation in GXe and LXe, an additional sharp single-PE feature at \SI{0.04}{\micro\second} was observed in MB0072, and, to a lesser extent, in MA0058. This feature potentially arises from He$^+$ ions present in the vacuum tubes. The persistence of this feature during dark current studies supports a non-LED origin. The significantly above-average abundance of the same relative PMT-to-PMT time difference between dark pulses in the paired PMTs further indicates afterpulse-induced light emission, as previously reported in XENONnT characterizations~\cite{antochiImprovedQualityTests2021}.

\begin{table}[!t]
    \centering
    \caption[Afterpulse rates of R12699-406-M4 PMTs in different environments.]{Measured afterpulse rates of five Hamamatsu R12699-406-M4 PMTs during MarmotX characterization at room temperature and in LXe. Rates are reported per LED-induced photoelectron (PE), including: all identified AP candidates, only those classified as separable, and separable APs with delay $> \SI{50}{ns}$ and area $> \SI{2}{PE}$ (see red boundaries in \autoref{fig:r12699_afterpulses}(b)). Measurements in Run~2 (LXe) were performed at lower occupancy, increasing uncertainties. PMTs MB0080 and MB0113 were excluded due to impaired pulse separability.}
    \label{tab:r12699_ap}
    
    \begin{tabular}{cl|ccc}
        \toprule
        & & \multicolumn{3}{c}{\textbf{Afterpulse Rate [\%/PE] --- Room Temperature}} \\
        & \textbf{PMT} & \textbf{All APs} & \textbf{Separable APs} & \textbf{Sep. APs} $>$2\,PE, $>$50\,ns \\
        \midrule
        \multirow{3}{*}{\rotatebox[origin=c]{90}{Run 1}} 
            & MA0055 & $0.94 \pm 0.08$ & $0.65 \pm 0.05$ & $0.223 \pm 0.018$ \\
            & MA0058 & $2.00 \pm 0.20$ & $1.06 \pm 0.12$ & $0.350 \pm 0.040$ \\
            & MB0072 & $1.15 \pm 0.06$ & $0.93 \pm 0.05$ & $0.307 \pm 0.016$ \\
        \midrule
        \multirow{2}{*}{\rotatebox[origin=c]{90}{Run 2}} 
            & MB0015 & $1.70 \pm 0.30$ & $1.40 \pm 0.20$ & $0.370 \pm 0.060$ \\
            & MB0090 & $1.37 \pm 0.11$ & $1.02 \pm 0.08$ & $0.350 \pm 0.030$ \\
        \midrule
        & \textbf{Mean} & $\mathbf{1.43 \pm 0.15}$ & $\mathbf{1.01 \pm 0.10}$ & $\mathbf{0.32 \pm 0.02}$ \\
        \bottomrule
        \addlinespace[1ex]
        \toprule
        & & \multicolumn{3}{c}{\textbf{Afterpulse Rate [\%/PE] --- LXe}} \\
        & \textbf{PMT} & \textbf{All APs} & \textbf{Separable APs} & \textbf{Sep. APs} $>$2\,PE, $>$50\,ns \\
        \midrule
        \multirow{3}{*}{\rotatebox[origin=c]{90}{Run 1}} 
            & MA0055 & $1.16 \pm 0.08$ & $0.84 \pm 0.06$ & $0.262 \pm 0.019$ \\
            & MA0058 & $1.30 \pm 0.20$ & $0.76 \pm 0.12$ & $0.230 \pm 0.040$ \\
            & MB0072 & $2.45 \pm 0.15$ & $2.18 \pm 0.13$ & $0.318 \pm 0.019$ \\
        \midrule
        \multirow{2}{*}{\rotatebox[origin=c]{90}{Run 2}} 
            & MB0015 & $1.07 \pm 0.16$ & $0.90 \pm 0.14$ & $0.290 \pm 0.050$ \\
            & MB0090 & $1.02 \pm 0.17$ & $0.76 \pm 0.13$ & $0.210 \pm 0.040$ \\
        \midrule
        & \textbf{Mean} & $\mathbf{1.40 \pm 0.20}$ & $\mathbf{1.10 \pm 0.20}$ & $\mathbf{0.262 \pm 0.016}$ \\
        \bottomrule
    \end{tabular}
\end{table}

\autoref{tab:r12699_ap} summarizes AP rates for different operating conditions. Average rates over a \SI{2}{\micro\second} window are \SI{1.43(15)}{\percent/PE} at room temperature and \SI{1.4(2)}{\percent/PE} in LXe. Applying the separability criterion reduces this to \SI{1.01(10)}{\percent/PE} (room temperature) and \SI{1.1(2)}{\percent/PE} (LXe). For APs with area > \SI{2}{PE} and delays > \SI{50}{ns}, mean rates are \SI{0.32(2)}{\percent/PE} and \SI{0.262(16)}{\percent/PE}, respectively.
No significant changes beyond $2\sigma$ were observed between room temperature and xenon operation, except for PMT MB0072, where the elevated AP rate is attributed to the unique fast component at \SI{0.04}{\micro\second}. 

For comparison, R11410-21 PMTs exhibit AP rates of \SI{1.4(12)}{\percent/PE} (room temperature) and \SI{8.6(22)}{\percent/PE} (LXe) over a \SI{4}{\micro\second} window~\cite{barrowQualificationTestsR11410212017}. However, these values are not directly comparable due to differing AP characteristics and separability. In R11410 PMTs, increased AP rates in LXe were dominated by a diffuse single-PE component, possibly from micro light emission, an effect not observed in the R12699-406-M4 devices tested here.

    \section{Application in a Time Projection Chamber} \label{sec:tpc}

To evaluate the performance of the \model PMT in a dual-phase TPC environment, two units were deployed in the XAMS dual-phase xenon TPC R\&D facility~\cite{HOGENBIRK201687}, located at Nikhef in Amsterdam. This section outlines the working principle of a dual-phase TPC and demonstrates the successful integration and functionality of the \model PMT in such a system. It furthermore shows how the separate anode readout of a single PMT unit can be used for lateral position reconstruction.

When a particle interacts within the liquid xenon target of a TPC, its energy deposition leads to excitation, ionization and heat production in the medium \cite{AprileDoke2010,Baudis:2023pzu}. Since the thermal energy is not detectable with this technology, it is neglected in the following. The excited atoms form short-lived dimer states (excimers), which de-excite by emitting \SI{175}{nm} UV photons~\cite{FUJII2015293}. This prompt light is known as the S1 signal. The ionization electrons are drifted toward the liquid-gas interface under the influence of a uniform electric field, referred to as the drift field. At the interface, a stronger extraction field between the gate and anode electrodes pulls the electrons into the gas phase, where they are further accelerated. These accelerated electrons excite xenon atoms through collisions, resulting in proportional scintillation light, known as the S2 or delayed signal. Both S1 and S2 signals are detected by photosensors placed at the top and the bottom of the TPC. The pattern of the S2 observed by the top photosensor array enables reconstruction of the particle interaction position in the transverse ($x$-$y$) plane. Simultaneously, the time difference between the S1 and S2 signals, given a known and constant electron drift velocity, provides the depth ($z$-coordinate) of the interaction. Together, these measurements allow full three-dimensional event reconstruction.


\subsection{The XAMS Facility: Setup, Data Processing and Selection}
\label{sec:xams}

\begin{figure}[!t]
	\sbox\twosubbox{%
		\resizebox{\dimexpr.99\textwidth-1em}{!}{%
			\includegraphics[trim={0 0 0 0},clip,height=3cm]{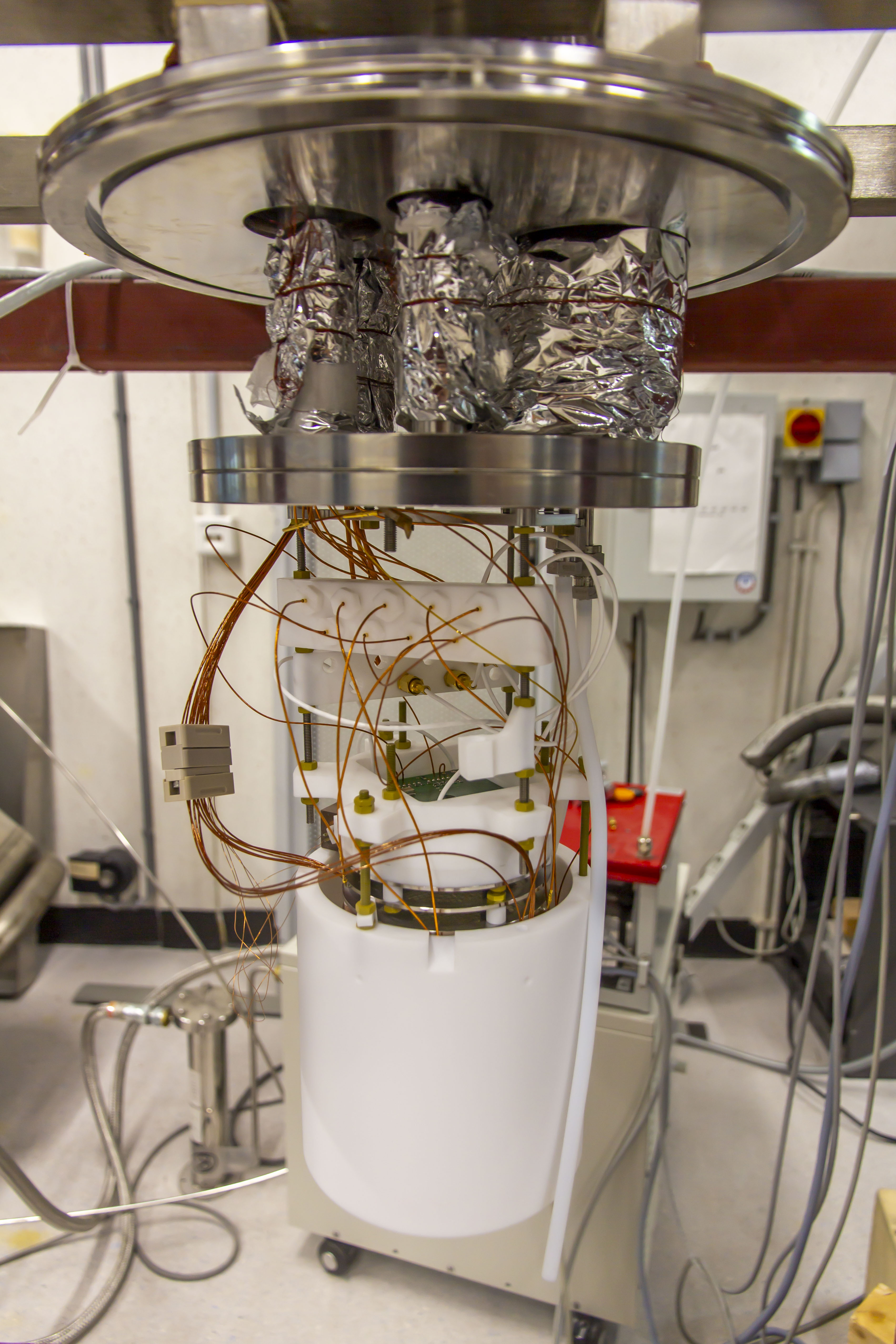}%
			\includegraphics[trim={10 0 10 0},clip,height=3cm]{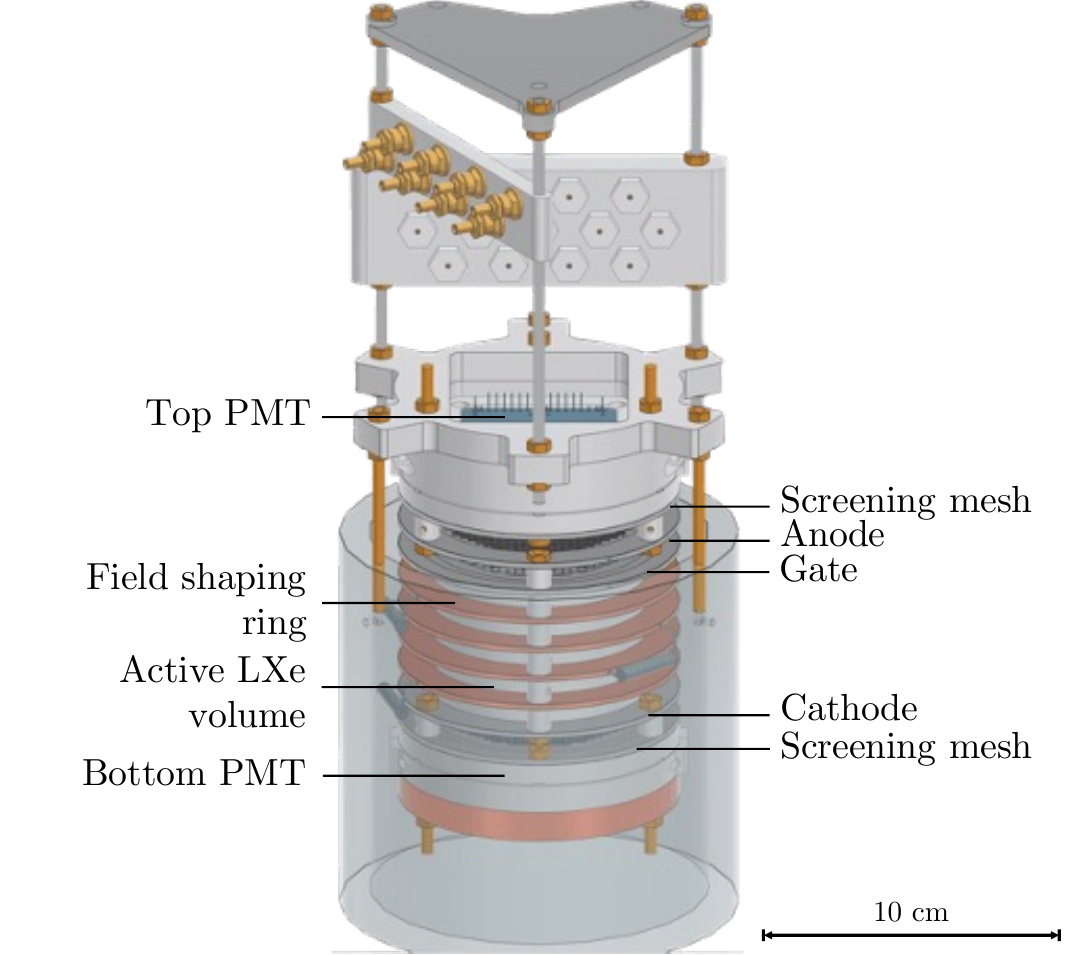}%
		}%
	}
	\setlength{\twosubht}{\ht\twosubbox}
	
	\centering	
	
	\subcaptionbox{}{%
		\includegraphics[trim={0 0 0 0},clip,height=0.99\twosubht]{figures/xams_photo.JPG}%
	}\quad
	\subcaptionbox{}{%
		\includegraphics[trim={10 0 10 0},clip,height=0.99\twosubht]{figures/xams_cad_annotated_10cm.pdf}%
	}
    \caption[The XAMS Time Projection Chamber]{The XAMS time projection chamber. \textbf{(a)}~Photograph of the assembled detector setup, featuring two \model PMTs installed at the top and bottom of the chamber. \textbf{(b)}~CAD rendering of the TPC, illustrating key structural components and the internal arrangement of the PMTs.} 

    \label{fig:xams_setup}
\end{figure}

The XAMS TPC, shown in \autoref{fig:xams_setup}, has an active, PTFE-lined volume with a height of \SI{5.1}{cm} and a diameter of \SI{6.2}{cm}, corresponding to \SI{154}{cm^3} or approximately \SI{445}{g} of LXe. Xenon is continuously circulated through a high-temperature SAES MonoTorr PS3-MT3-R-2 getter to remove electronegative impurities. A drift field of \SI{100}{V/cm} is applied between the cathode at \SI{-1.7}{kV} and the gate mesh at \SI{-1.2}{kV}, with field uniformity maintained by five copper field-shaping rings. The extraction region is located between the gate and anode mesh, with a separation of \SI{8}{\milli\meter}. An extraction voltage of \SI{3.2}{kV} is applied between the two electrodes, where the liquid-gas interface is located \SI{3}{mm} above the gate. The S1 and S2 signals are detected by two R12699-406-M4 PMTs, one positioned at the top and one at the bottom of the TPC. Screening meshes, biased at \SI{-0.5}{kV} (top) and \SI{-1}{kV} (bottom), shield the PMTs from the electric fields in the active volume. The four anode segments of the top PMT are read out independently, enabling lateral position reconstruction, while the bottom PMT outputs are combined and read out as a single channel (see \autoref{sec:darwin_2inch_readout}). The top and bottom PMTs were operated at bias voltages of \SI{-1000}{V} and \SI{-1080}{V}, respectively, with the latter exceeding the nominal voltage to enhance the gain. An optical fiber installed in the TPC delivers LED light pulses for gain calibration of each channel. XAMS was operated over five data-taking periods, each lasting approximately two weeks, over a total span of thirteen months. During each period, gain stability was maintained within \SI{\ca 1}{\percent} for the top PMT channels and within \SI{\ca 5}{\percent} for the bottom channel.

\begin{figure}[!t]
	\sbox\twosubbox{%
		\resizebox{\dimexpr.99\textwidth-1em}{!}{%
			\includegraphics[trim={7 5 5 5},clip,height=3cm]{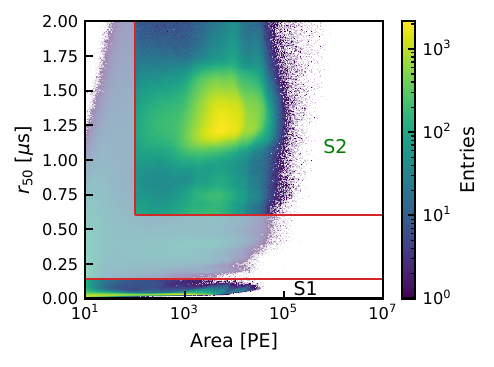}%
			\includegraphics[trim={5 5 7 5},clip,height=3cm]{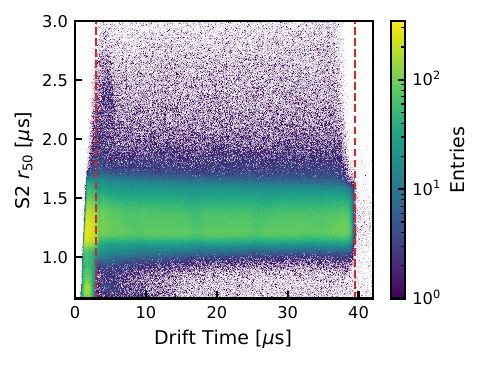}%
		}%
	}
	\setlength{\twosubht}{\ht\twosubbox}
	
	\centering	
	
	\subcaptionbox{}{%
		\includegraphics[trim={7 5 5 5},clip,height=0.99\twosubht]{figures/006734_006762_s1s2_highlight_rasterized.pdf}%
	}\quad
	\subcaptionbox{}{%
		\includegraphics[trim={5 5 7 5},clip,height=0.99\twosubht]{figures/006734_006762_drifttime_s2width_rasterized.pdf}%
	}
  \caption[XAMS data classification and selection]{Data classification and event selection in XAMS, using an external $^{22}$Na calibration source. \textbf{(a)}~Classification of S1 and S2 signals based on pulse area and the time interval containing \SI{50}{\percent} of the total area ($r_{50}$). Typical S1 signals are short, while S2 signals appear significantly broader due to electroluminescence in the extraction region. Waveforms that have a width between 0.14 and \SI{0.60}{\micro\second} are mostly S2 signals generated from interactions in the small liquid layer above the gate mesh. In this region, the strong extraction field causes the more narrow spread in electron arrival time, thus, $r_{50}$ compared to the interactions in the drift region. Waveforms with areas below \SI{100}{PE} and widths above \SI{0.14}{\micro\second} are mostly S2 signals generated from interactions in the gas gap. Interactions from these two regions, thus their corresponding S2 signals, are not considered, as indicated by the shaded area. \textbf{(b)}~Distribution of events as a function of drift time. The rise in event count below \SI{3}{\micro\second} corresponds to interactions above the gate electrode, while the cutoff at \SI{39.5}{\micro\second} indicates the cathode position, beyond which no events are expected. For position reconstruction, only events with drift times between \SI{3}{\micro\second} and \SI{39.5}{\micro\second} are considered.}
\label{fig:data_classification_selection}

\end{figure}

Data processing was performed using \texttt{amstrax}~\cite{angevaare_amstrax}, an open-source data analysis framework  built on \texttt{strax}~\cite{aalbers_strax}. \autoref{fig:data_classification_selection}~(a) illustrates the classification of S1 and S2 waveforms based on their area and time width ($r_{50}$), defined as the duration over which the pulse contains \SI{50}{\percent} of its area. Typical S1 pulses in XAMS have widths of approximately \SI{30}{ns}, set by the decay time of the recombined dimer states. Waveforms are classified as S1 signals if $r_{50} < \SI{0.14}{\micro\second}$. In contrast, S2 waveforms are broader, typically on the order of $\mathcal{O}$(\SI{}{\micro \second}), primarily due to the size of the gas gap where electroluminescence occurs. Electron diffusion contributes minimally in XAMS due to its short \SI{5}{cm} drift length. A waveform is classified as S2 if it has a width of at least \SI{0.60}{\micro\second} and a minimum area of 100\,PE. To identify physical S1-S2 pairs, each \textit{triggering waveform} (i.e., an S2) is grouped with nearby waveforms within a \SI{50}{\micro \second} window before and after the trigger. Within each event window, the largest S1 and S2 pulses are designated as the main interaction signals. Overlapping event windows are merged to ensure consistency.

The data presented here were acquired using an external $^{22}$Na calibration source. $^{22}$Na is a $\beta^+$ emitter that decays to an excited state of $^{22}$Ne, followed by positron annihilation into two \SI{511}{keV} back-to-back gamma rays, and a subsequent \SI{1274}{keV} gamma from de-excitation. Additional interactions originate from ambient background sources, including detector materials, laboratory radioactivity, and atmospheric muons. \autoref{fig:data_classification_selection}~(b) shows the distribution of S2 $r_{50}$ values as a function of drift time for a \SI{15}{hour} calibration dataset. For the position reconstruction, only events with drift times between \SI{3.5}{\micro \second} and \SI{39.5}{\micro \second} are selected, corresponding to interactions within the TPC’s active region, between the cathode and the gate mesh, spanning a vertical distance of \SI{5}{cm}.

\subsection{Lateral Position Reconstruction}

\begin{figure}[!t]
	\sbox\twosubbox{%
		\resizebox{\dimexpr.99\textwidth-1em}{!}{%
			\includegraphics[trim={5 0 0 0},clip,height=3cm]{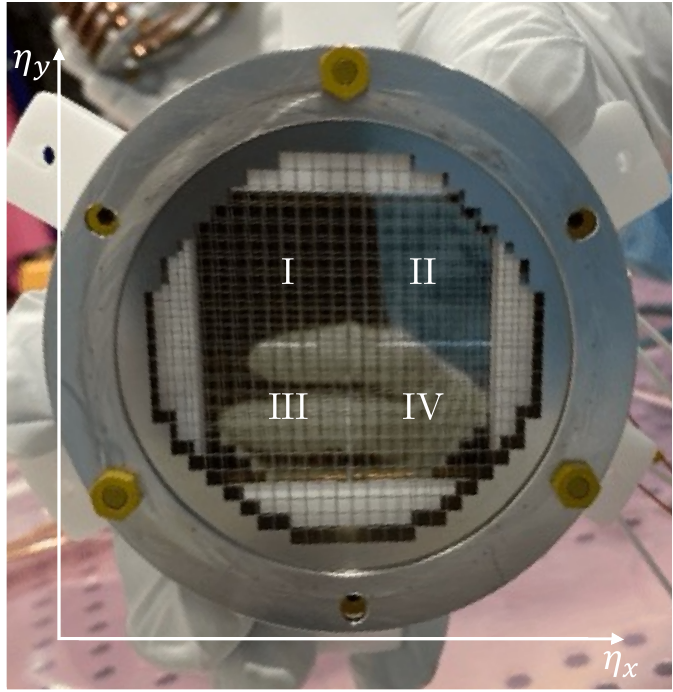}%
			\includegraphics[trim={0 0 10 0},clip,height=3cm]{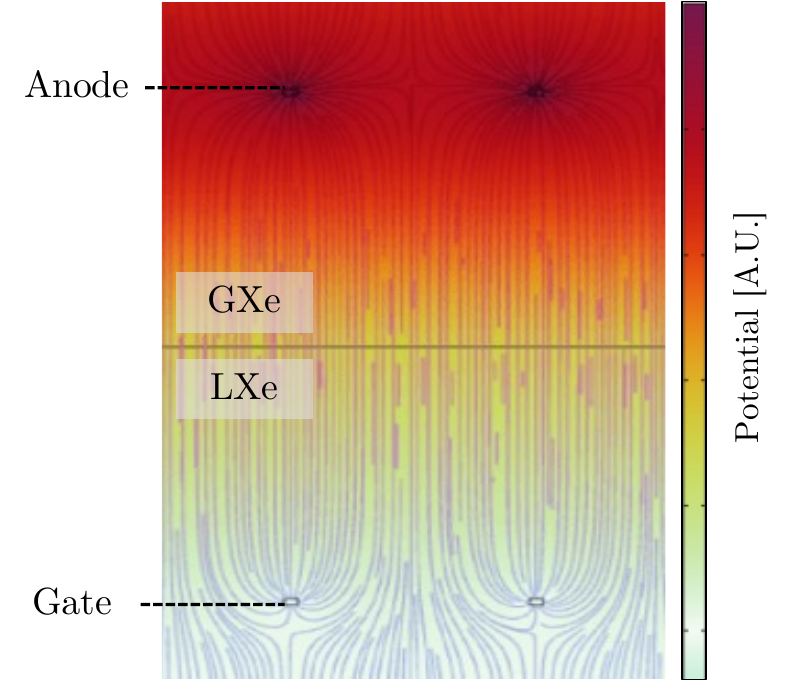}%
		}%
	}
	\setlength{\twosubht}{\ht\twosubbox}
	
	\centering	
	
	\subcaptionbox{}{%
		\includegraphics[trim={5 0 0 0},clip,height=0.99\twosubht]{figures/xams_top_PMT_holder_annotated.pdf}%
	}\quad
	\subcaptionbox{}{%
		\includegraphics[trim={0 0 10 0},clip,height=0.99\twosubht]{figures/electric_fieldlines_xams_A.U..pdf}%
	}
	\caption[Position reconstruction method and uncorrected results]{\textbf{(a)}~Photograph of the top PMT holder in XAMS, showing the mounted top screening mesh and the anode mesh positioned in front of the PMT. The coordinate axes and the four labeled PMT sectors correspond to the charge ratio coordinates defined in \autoref{eq:charge_ratio}. \textbf{(b)}~Simulation of the electric field near the gas–liquid xenon interface in XAMS, produced with \texttt{COMSOL}. The color scale indicates the electric potential, while the gray line between the anode and gate meshes marks the gas–liquid interface. Electrons originating from below the gate are guided upward between the gate wires toward the anode.}
\label{fig:pos_rec_method}

\end{figure}

Lateral position reconstruction in XAMS is based on the centroid method using \textit{charge ratios} (CRs), which describe the relative amount of light detected in each sector  of the top PMT. For each event, the charge ratio coordinates $\eta_{\text{x}}$ and $\eta_{\text{y}}$ are computed as 

\begin{equation}
\label{eq:charge_ratio}
    \eta_{\text{x}} = \frac{Q_\text{II}+Q_{\text{IV}}}{Q_\text{total}} \quad \textrm{and} \quad \eta_{\text{y}} = \frac{Q_\text{I}+Q_{\text{II}}}{Q_\text{total}},
\end{equation}

\noindent where $Q$ denotes the integrated signal (in PE) in each of the four PMT sectors (I-IV) and $Q_\text{total}$ is their sum. \autoref{fig:pos_rec_method}~(a) shows the top PMT mounted in its PTFE holder, with the top screening mesh and the anode mesh stacked in front. The four PMT quadrants are labeled according to the charge ratio definition.
\begin{figure}[t]
	\sbox\twosubbox{%
		\resizebox{\dimexpr.99\textwidth-1em}{!}{%
			\includegraphics[trim={7 5 5 5},clip,height=3cm]{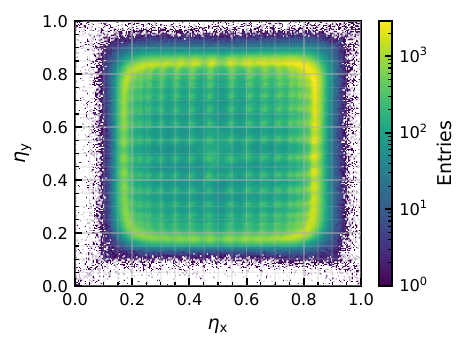}%
			\includegraphics[trim={5 5 7 5},clip,height=3cm]{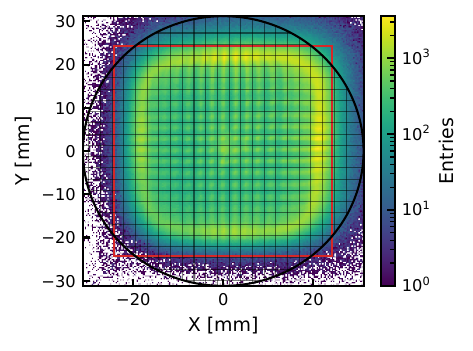}%
		}%
	}
	\setlength{\twosubht}{\ht\twosubbox}
	
	\centering	
	
	\subcaptionbox{}{%
		\includegraphics[trim={7 5 5 5},clip,height=0.99\twosubht]{figures/na_event_positions_uncorrected_runs_006734_006762_corrections_version_v1_II.pdf}%
	}\quad
	\subcaptionbox{}{%
		\includegraphics[trim={5 5 7 5},clip,height=0.99\twosubht]{figures/na_event_positions_corrected_runs_006734_006762_corrections_version_v1_PMT_PC.pdf}%
	}
	\caption[Position reconstruction results]{Results of the \textit{charge ratio}–based position reconstruction in XAMS. \textbf{(a)}~Reconstructed event positions before applying the grid correction. \textbf{(b)}~Reconstructed positions after grid correction. The positions of the gate grid wires are overlaid in black, and the red outline marks the boundary of the PMT photocathode. The grid wire pitch is \SI{2.6}{mm}.}
\label{fig:pos_rec_results}
\end{figure}
\autoref{fig:pos_rec_method}~(b) shows a simplified 2D simulation of the electric field in the extraction region, computed using \texttt{COMSOL}. The field lines indicate that electrons from below the gate are focused toward the midpoint between gate wires. The electroluminescence (S2) signals are generated subsequently at the same lateral position in the gas phase.

\begin{figure}[!t]
	\sbox\twosubbox{%
		\resizebox{\dimexpr.99\textwidth-1em}{!}{%
			\includegraphics[trim={7 5 5 5},clip,height=3cm]{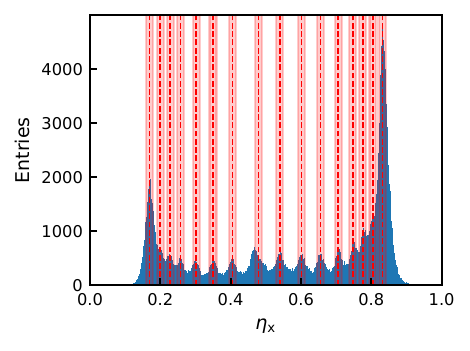}%
			\includegraphics[trim={5 5 7 5},clip,height=3cm]{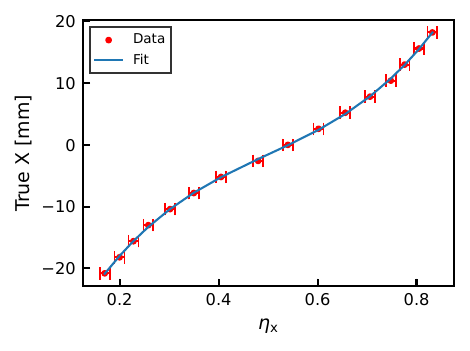}%
		}%
	}
	\setlength{\twosubht}{\ht\twosubbox}
	
	\centering	
	
	\subcaptionbox{}{%
		\includegraphics[trim={7 5 5 5},clip,height=0.99\twosubht]{figures/na_event_eta_CR_runs_006734_006762_corrections_version_v1.pdf}%
	}\quad
	\subcaptionbox{}{%
		\includegraphics[trim={5 5 7 5},clip,height=0.99\twosubht]{figures/na_peak_locations_correction_curve_runs_006734_006762_corrections_version_v1.pdf}%
	}

    \caption[Position reconstruction correction method]{Position reconstruction correction method based on known grid wire spacing. \textbf{(a)}~Extraction of dot positions for a constant slice in the vertical charge ratio coordinate, $0.47 < \eta_{\text{y}} < 0.49$. \textbf{(b)}~Derivation of the correction function: a third-degree polynomial is fitted to the measured $\eta_{\text{x}}$ values and their corresponding true positions.}
\label{fig:correction_method}
\end{figure}

The result of applying the charge ratio method to S2 signals from $^{22}$Na and background events is shown in \autoref{fig:pos_rec_results}~(a). A regular dot pattern appears, consistent with the focusing effect of the extraction field. A higher event density in the upper-right corner corresponds to the location of the external calibration source. The measured distance between the gate grid wires provides an absolute scale for correcting the positions from the charge ratios to true coordinates of the S2 light generation. \autoref{fig:correction_method}~(a) shows the distribution of dot positions in $\eta_{\text{x}}$ within a representative slice $0.47 < \eta_{\text{y}} < 0.49$. These reconstructed positions are then mapped to their known physical positions $X$ based on the gate grid pitch. As shown in \autoref{fig:correction_method}~(b), a third-degree polynomial is fit to define the correction function. The procedure is repeated for the $\eta_{\text{y}}$ coordinate using a horizontal slice, $0.47 < \eta_{\text{x}} < 0.49$, to determine the correction function for the $Y$ position. 

The distribution of corrected event positions is shown in \autoref{fig:pos_rec_results}~(b), with grid wire locations overlaid in black. The curved edges of the distribution are consistent with the partial optical shielding from the mesh frame, visible in \autoref{fig:pos_rec_method}~(a). The geometrical alignment between the shaded PMT regions and the low-density regions in the corrected map demonstrates the effectiveness of the reconstruction. 
Although a detailed optical model or calibration measurement of the PMT response is not available for our application, qualitative conclusions about position resolution can still be drawn. The reconstruction resolution degrades towards the TPC edges, as evidenced by the  merging of four grid wire-induced dots in those regions. The spatial resolution is inversely proportional to the slope of the correction function shown in \autoref{fig:correction_method}. In the central region ($-20~\text{mm} < X,Y < 20~\text{mm}$) within the LXe target (radius of \SI{31}{mm}), dot separation remains distinct. The grid wire pitch of \SI{2.6}{mm} thus provides a conservative upper bound on the resolution in this region. A degradation in resolution near the edges is expected and consistent with known limitations of TPC-based position reconstruction \cite{xenoncollaboration2024xenonntanalysissignalreconstruction}. These effects arise from reduced photon statistics near inactive regions and increased reflections from PTFE surfaces. Nonetheless, this study demonstrates that lateral position reconstruction is achievable in a dual-phase xenon TPC using a single segmented PMT.

    \section{Summary and Outlook} 
\label{sec:discussion_conclusion}

We have performed a detailed characterization of several \model PMTs at room temperature and, for the first time, in cryogenic xenon environments. The results indicate promising performance characteristics that support the applicability of this sensor model in future LXe TPC experiments. All investigated PMTs demonstrated gains exceeding \SI{2e6}{} at the nominal operating voltage, with gain stability observed within \SI{3.6}{\percent} over a period of 62~days in liquid xenon and no indication of gain drift. Dark count rates, normalized  by photocathode area, were found to be reduced by a factor of $3 \pm 2$ compared to the widely used 3-inch R11410 PMTs. However, the afterpulsing behavior of the \model~PMT warrants special attention due to its significantly faster timing, which can lead to overlap with the primary light-induced signal. A summary of the key performance metrics for the \model PMT, along with  a comparison to the R11410 model, is provided in \autoref{tab:pmts_comparioson}. 

The results are consistent with the performance of this PMT model in a cold nitrogen environment, as reported in \cite{Yun:2024oxp}. The reported SPE resolution of \SI{42}{\percent} in the study in cold nitrogen gas was extracted at a single nominal bias voltage, using a different voltage divider circuit optimized to increase gain. This study reports the SPE-resolution for a fixed gain of \num{\ca 2e6}, enabled by measurements at different bias voltages. A comparison of the average SPE resolutions is therefore not trivial. Another distinction between the studies is that for this work, a PMT configuration was chosen that favors the ability to monitor potential light emission from the PMTs themselves, at the cost of a non-uniform LED illumination of the photocathodes. This potentially results in an underestimation of the SPE resolution, as position-dependent effects on the gain may influence the estimates.

The successful deployment and operation of two \model units in a compact dual-phase xenon TPC demonstrated the feasibility of this PMT type in its intended application. In addition to stable operation, the sensor enabled lateral position reconstruction based on the segmented $2 \times 2$ multi-anode readout, confirming the spatial resolution potential for LXe-based detectors. 

\begin{table}[t]
    \centering
    \caption[Comparison R12699-406-M4 and R11410-21 PMTs.]{Comparison of the \model~PMT characterized in this work with the R11410-21 model used in the XENON1T/nT experiments. All values refer to measurements in cryogenic xenon environments and nominal operation voltages, unless stated otherwise. Comparability of certain parameters, especially those affected by limited afterpulse separability, is discussed in the main text. A radiopurity comparison is provided in \autoref{tab:gator_results_r12699}.} 
    \begin{tabular}{l|cc} 
        \toprule
        & \textbf{R12699-406-M4} & \textbf{R11410-21}\\
        \midrule
        Dimensions (longitudinal / lateral) & \SI{14.8}{mm} / \SI{56}{mm}~\cite{hamamatsuphotonicsk.k.FlatPanelType2020} & \SI{114}{mm} / \SI{77.5}{mm}~\cite{hamamatsuphotonicsk.k.PhotomultiplierTubeR11410212014} \\ 
        Packing density & \SI{75.0}{\percent} & \SI{61.8}{\percent} \\ 
        Dynode structure & Metal channel~\cite{hamamatsuphotonicsk.k.FlatPanelType2020} & Box \& linear-focused~\cite{hamamatsuphotonicsk.k.PhotomultiplierTubeR11410212014} \\
        Number of dynode stages & 10~\cite{hamamatsuphotonicsk.k.FlatPanelType2020} & 12~\cite{hamamatsuphotonicsk.k.PhotomultiplierTubeR11410212014} \\
        Quantum efficiency at \SI{175}{nm} & \SI{33}{\percent}~\cite{hamamatsuphotonicsk.k.FlatPanelType2020} & \SI{32.5}{\percent}~\cite{hamamatsuphotonicsk.k.PhotomultiplierTubeR11410212014} \\
        Operation voltage (nominal / max.) & \SI{1000}{V} / \SI{1100}{V}~\cite{hamamatsuphotonicsk.k.FlatPanelType2020} & \SI{1500}{V} / \SI{1750}{V}~\cite{hamamatsuphotonicsk.k.PhotomultiplierTubeR11410212014} \\ 
        Gain ($\mathrm{T}_{\mathrm{room}}$) & \SI{3.3(7)e6}{} & \SI{8.4(23)e6}{}~\cite{antochiImprovedQualityTests2021} \\ 
        SPE resolution (gain $\sim \SI{2e6}{}$, $\mathrm{T}_{\mathrm{room}}$) & \SI{37(2)}{\percent} & \SI{30(3)}{\percent}~\cite{barrowQualificationTestsR11410212017} \\ 
        Dark count rate ($> \nicefrac{1}{4}$\;PE) & \SI{0.4(2)}{Hz/cm^2} & \SI{1.4(7)}{Hz/cm^2}~\cite{barrowQualificationTestsR11410212017} \\ 
        Time response (RT / transit time / TTS) & 1.2 / 5.9 / \SI{0.41}{ns}~\cite{hamamatsuphotonicsk.k.FlatPanelType2020} & 5.5 / 46 / \SI{9}{ns}~\cite{hamamatsuphotonicsk.k.PhotomultiplierTubeR11410212014} \\
        Separable afterpulse rate & \SI{1.1(2)}{\percent/PE} & \SI{8.6(22)}{\percent/PE}~\cite{barrowQualificationTestsR11410212017} \\ 
        \bottomrule 
    \end{tabular}
    \label{tab:pmts_comparioson}		
\end{table}

Scaling to a large TPC would require a photosensor density of approximately \SI{300}{PMTs/m^2}, making compact, low-radioactivity designs such as the \model PMT particularly attractive.  Future studies will explore extended capabilities, including pulse-shape discrimination using a planned small-scale dual-phase TPC with a $2 \times 4$ array of \model PMTs to be operated in the MarmotX facility.

\appendix
\newpage
\section{Signal Readout}
\label{sec:darwin_2inch_readout}

\begin{figure}[!p]
\centering
\includegraphics[width=1.3\linewidth, trim={80 200 60 100}, clip, angle=90, origin=c]{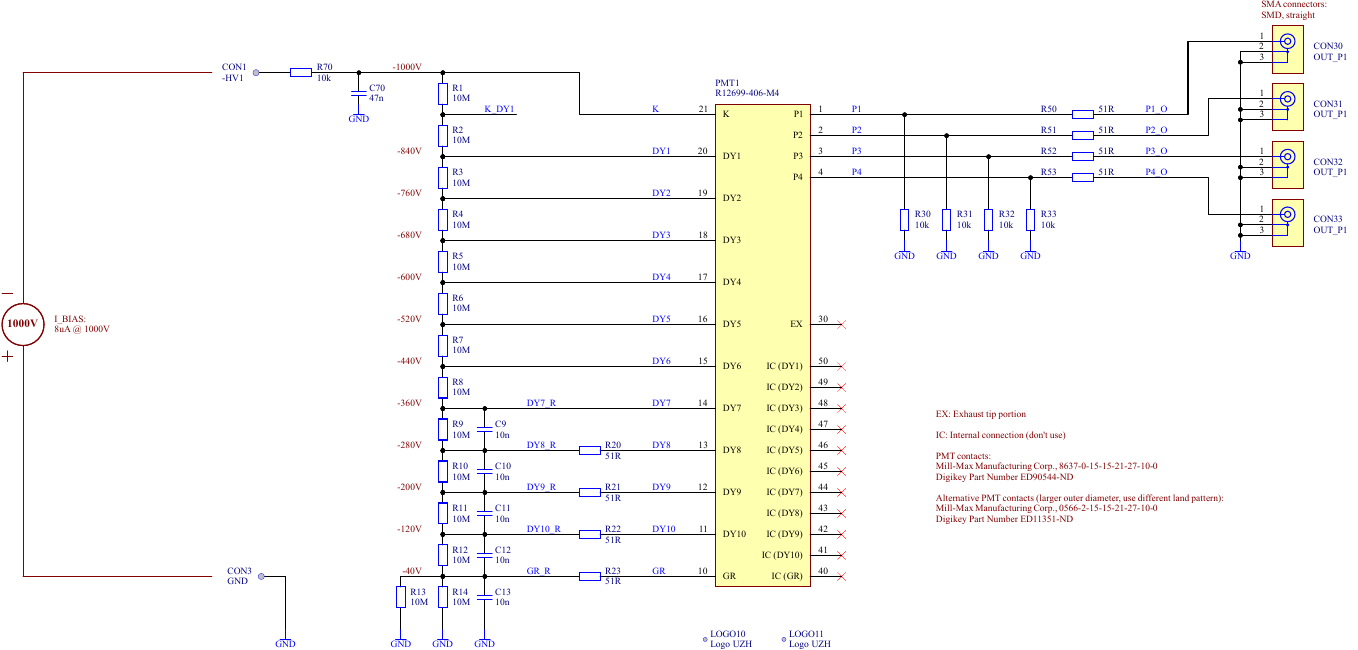}
\vspace*{0.2cm}
\caption[Schematic of the Hamamatsu R12699-406-M4 PMT base.]{Schematic of the Hamamatsu R12699-406-M4 PMT voltage divider for separate anode readout. The configuration for combined anode readout is obtained by setting R31--R33 to \texttt{OPEN} and combining R50--R53 to a single \texttt{51R} with subsequent SMA connector. Image credit: Daniel Florin, UZH.}
\label{fig:r12699_pmt_base_schematic}
\end{figure}

A crucial factor in PMT performance and signal quality is the voltage divider circuit, which defines the voltage distribution across the electron amplification stages. The circuit used in this work is shown in \autoref{fig:r12699_pmt_base_schematic}. In later development stages, SMA connectors replaced directly soldered signal cables to reduce the PCB footprint and support denser array configurations. The design also transitioned from through-hole technology (THT) to surface-mount devices (SMD) to minimize the risk of sparking between the negatively biased PMT body and exposed leads.

The circuit is operated in a negative bias configuration using a single high-voltage source. The total resistance of $\SI{125}{M\Omega}$ results in a base current of approximately \SI{8}{\micro\ampere} at \SI{-1.0}{kV}, with equal voltage division across dynodes, except for a doubled resistance between the photocathode and first dynode to enhance collection efficiency, and a halved resistance between the final dynode and the anode. Auxiliary components improve signal quality. A low-pass filter ($f_c = \SI{0.34}{kHz}$) on the HV line suppresses noise. To mitigate ring-down effects from the PMT and readout system reactive elements, four $\SI{51}{\Omega}$ damping resistors are added to the final dynodes. Additionally, five \SI{10}{nF} decoupling capacitors stabilize voltages during strong illumination by supplying extra current to the final stages.

During the MarmotX studies (\autoref{sec:characterization}), all four anodes were combined and read out through a $\SI{51}{\Omega}$ output. Anode trace lengths were matched on the PCB to avoid timing skew. A slightly modified version for separate anode readout was used in the XAMS TPC setup (\autoref{sec:tpc}).
\newpage
\section{Radiopurity Optimization}
\label{sec:gator}

Radiopurity is critical for the application of \model PMTs in future low-background experiments such as the proposed XLZD and PandaX-xT. The optimization carried out by the manufacturer over the past years was informed by material radioassay with the Gator low-background germanium spectrometer~\cite{araujoUpgradedLowbackgroundGermanium2022c, bismarkTestsFundamentalsQuantum2024}. 

\begin{sidewaystable}
\centering
\caption[Gator radioassay results for R12699-406-M4 PMTs.]{Measured activities of radioactive isotopes in various R12699-406-M4 PMT development stages, obtained using the Gator low-background facility. Values are normalized to sensitive photocathode area. Upper limits are quoted at 90\% CL. For comparison, data for legacy R8520-06 and R11410-21 PMTs are also shown.}
\label{tab:gator_results_r12699}

\begin{tabular*}{\textwidth}{@{\extracolsep{\fill}} r|lcccl}
    \toprule
    \textbf{ } & \textbf{ } & \textbf{Mass} & \textbf{Livetime Sample /} & \textbf{ } & \textbf{ } \\
    \textbf{ID} & \textbf{Sample Description} & \textbf{[g/PMT]} & \textbf{Background [d]} & \textbf{Units} & \textbf{Notes} \\
    \midrule
    \multirow{2}{*}{0} & Initial model (2× R12699-406-M4) & \multirow{2}{*}{$2\times104$} & \multirow{2}{*}{41.7 / 73.7} & \multirow{2}{*}{mBq/cm$^2$} & Full PMT volume \\
     & & & & & Glass bead region only \\
    \multirow{2}{*}{1} & Improved LRI metal (2× R12699-406-M4/NG) & \multirow{2}{*}{$2\times104$} & \multirow{2}{*}{47.8 / 73.7} & \multirow{2}{*}{mBq/cm$^2$} & Full PMT volume \\
     & & & & & Glass bead region only \\
    \multirow{2}{*}{2} & Improved LRI glass (3× R12699-406-M4/NG) & \multirow{2}{*}{$3\times104$} & \multirow{2}{*}{40.8 / 32.2} & \multirow{2}{*}{mBq/cm$^2$} & Full PMT volume \\
     & & & & & Glass bead region only \\
    \midrule
    3 & R8520-06-Al (XENON100) & Various & Various & mBq/cm$^2$ & From~\cite{aprileMaterialScreeningSelection2011} \\
    4 & R11410-21 (XENONnT) & Various & Various & mBq/cm$^2$ & From~\cite{aprileMaterialRadiopurityControl2022} \\
    \bottomrule
\end{tabular*}

\vspace{1em}

\begin{tabular*}{\textwidth}{@{\extracolsep{\fill}} r|cccccccccc}
    \toprule
    \textbf{ID} & $^{238}$U & $^{226}$Ra & $^{228}$Ra & $^{228}$Th & $^{235}$U & $^{60}$Co & $^{40}$K & $^{137}$Cs & $^{54}$Mn & $^{58}$Co \\
    \midrule
    \multirow{2}{*}{0} & $<0.260$ & $0.026(4)$ & $<0.028$ & $<0.023$ & $<0.008$ & $0.055(5)$ & $1.47(16)$ & $<0.005$ & $<0.006$ & $<0.005$ \\
                       & $<0.376$ & $0.028(4)$ & $<0.031$ & $<0.023$ & $<0.011$ & $0.064(6)$ & $1.78(19)$ & $<0.006$ & $<0.007$ & $<0.006$ \\
    \multirow{2}{*}{1} & $<0.294$ & $0.023(4)$ & $<0.023$ & $<0.015$ & $<0.009$ & $0.0036(16)$ & $1.45(15)$ & $<0.005$ & $0.0070(15)$ & $0.0103(18)$ \\
                       & $<0.425$ & $0.026(4)$ & $<0.026$ & $<0.018$ & $<0.012$ & $0.0041(18)$ & $1.76(19)$ & $<0.006$ & $0.0086(19)$ & $0.012(2)$ \\
    \multirow{2}{*}{2} & $<0.404$ & $<0.015$ & $<0.022$ & $<0.017$ & $<0.010$ & $0.0057(17)$ & $1.29(14)$ & $<0.006$ & $0.0068(17)$ & $0.0055(15)$ \\
                       & $<0.621$ & $<0.017$ & $<0.025$ & $<0.020$ & $<0.013$ & $0.007(2)$   & $1.57(17)$ & $<0.008$ & $0.008(2)$   & $0.0067(18)$ \\
    \midrule
    3 & $<3.569$ & $<0.067$ & $<0.140$ & $0.071(17)$ & $<0.159$ & $0.144(10)$ & $2.86(18)$ & $<0.024$ & -- & -- \\
    4 & $0.28(6)$ & $0.0146(6)$ & $0.015(2)$ & $0.0143(6)$ & $0.012(3)$ & $0.0326(9)$ & $0.441(16)$ & $<0.004$ & -- & -- \\
    \bottomrule
\end{tabular*}
\end{sidewaystable}

Initial radioassay measurements were conducted on two commercially available R12699-406-M4 units (MA0055 and MA0058). \Autoref{tab:gator_results_r12699} compares the radioassay values to previous Gator screening results for the optimized R11410-21 units for XENONnT~\cite{aprileMaterialRadiopurityControl2022} and Hamamatsu R8520-06 model~\cite{aprileMaterialScreeningSelection2011}. Activities  are normalized to  photocathode area (R12699 \SI{\ca 23.52}{cm^2}, R11410 \SI{\ca 32.2}{cm^2}, R8520 \SI{\ca 4.22}{cm^2}) to allow comparison. Where above-threshold detections were made, the R12699 PMT showed 2--3 times higher activities for the isotopes  $^{226}$Ra, $^{60}$Co, and $^{40}$K compared to the R11410. However, it outperformed the older R8520 model. This may be attributed to its lower material content per photocathode area. For isotopes with only upper limits, no conclusive comparison is possible.

\autoref{tab:gator_results_r12699} includes estimates based on two scenarios for radionuclide distribution: homogeneous and localized in the glass beads near the leads, the latter assumption being supported by prior studies~\cite{aprileLoweringRadioactivityPhotomultiplier2015, Yun:2024oxp}. Three mechanical sample PMTs  with improved low-radioactivity (LRI) metal components were screened. These showed similar $^{226}$Ra and $^{40}$K levels as the initial model, but slightly elevated $^{58}$Co and $^{54}$Mn, likely due to extended transport and exposure to cosmic rays before screening. Notably, the $^{60}$Co activity was reduced by \SI{93(3)}{\percent}, achieving values an order of magnitude lower than those of the R11410. Further improvements were made with samples using LRI glass in the bead regions. These showed reduced $^{58}$Co and $^{226}$Ra activities, with levels on par with or below those of the R11410.

According to the manufacturer, our Gator results are broadly consistent with their estimates from the individual materials, except for $^{40}$K, where measurements indicate roughly ten times higher activity. This discrepancy, also observed in R11410 screenings~\cite{piastraMaterialsRadioassayXENON1T2017}, may stem from the bialkali photocathode deposition process. A comprehensive material assay for the R11410-21 and its raw components can be found in~\cite{Yun:2024oxp}.

\newpage
\section{Dark Count Origins}
\label{sec:dc_populations}

\begin{figure}[!t]
	\centering
	
	\sbox\twosubbox{%
		\resizebox{\dimexpr.99\textwidth-1em}{!}{%
			\includegraphics[trim={7 0 0 0},clip,height=3cm]{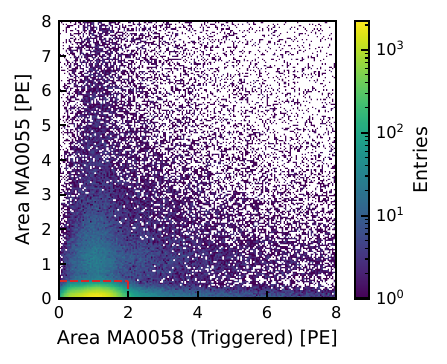}%
			\includegraphics[trim={0 0 7 0},clip,height=3cm]{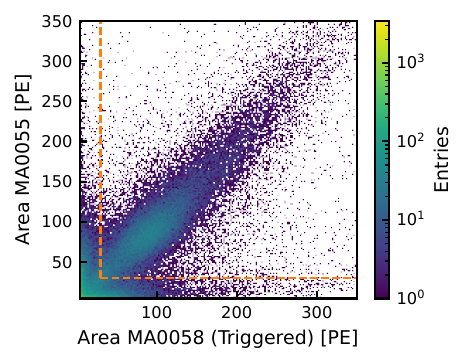}%
		}%
	}
	\setlength{\twosubht}{\ht\twosubbox}
	
	\subcaptionbox{}{%
		\includegraphics[trim={7 0 0 0},clip,height=\twosubht]{figures/areas2dlow_AP_ch0-1_disctrig_trigch0_U0-3_-1000V_20230426_0}%
	}\quad
	\subcaptionbox{}{%
		\includegraphics[trim={0 0 7 0},clip,height=\twosubht]{figures/areas2d_AP_ch0-1_disctrig_trigch0_U0-3_-1000V_20230426_0}%
	}
	
	\sbox\twosubbox{%
		\resizebox{\dimexpr.99\textwidth-1em}{!}{%
			\includegraphics[trim={7 0 0 0},clip,height=3cm]{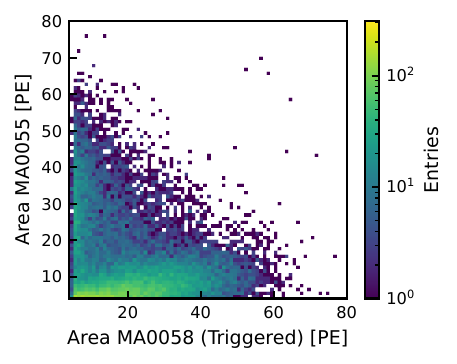}%
			\includegraphics[trim={0 0 7 0},clip,height=3cm]{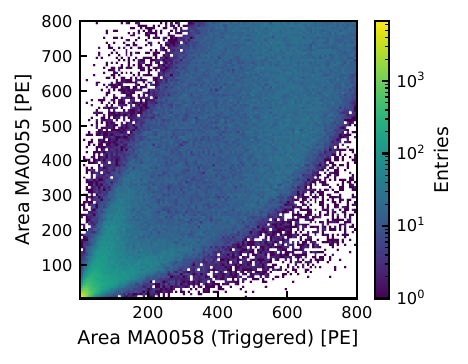}%
		}%
	}
	\setlength{\twosubht}{\ht\twosubbox}
	
	\subcaptionbox{}{%
		\includegraphics[trim={7 0 0 0},clip,height=\twosubht]{figures/areas2dmed_SPE_MA0055_MA0058_basev103_U0-1000V_U1-1000V_channeltrigger-ch0_Co60below_20221019_00}%
	}\quad
	\subcaptionbox{}{%
		\includegraphics[trim={0 0 7 0},clip,height=\twosubht]{figures/areas2d_AP_ch0-1_disctrig_trigch0_U0+1+3_-1000V_20230508_1}%
	}
	
	\caption[Dark count populations R12699-406-M4 PMT.]{Dark count populations identified via area correlation between the triggered PMT MA0058 and its paired unit MA0055. \textbf{(a)}~Low-PE region at room temperature showing dominant uncorrelated thermionic emission (bottom-left cluster). \textbf{(b)}~High-PE region highlighting strongly correlated events around 100\,PE, consistent with cosmic muons. \textbf{(c)}~Response to a collimated $^{60}$Co source placed below the PMT pair, showing asymmetry indicative of gamma-induced activity. \textbf{(d)}~High-amplitude correlated events from xenon scintillation with both PMTs submerged in LXe.}\label{fig:r12699_dc_polulations}
\end{figure}

To study correlations between paired PMTs and disentangle the sources of dark counts, the discriminator trigger from one PMT was used to initiate the ADC readout for both PMTs in a pair. \autoref{fig:r12699_dc_polulations} illustrates correlated area histograms between PMT MA0058 (triggered) and its counterpart MA0055. 

Panel~(a) highlights the dominant population of low-area signals ($<\SI{2}{PE}$ in MA0058 and $<\SI{0.5}{PE}$ in MA0055), attributed to uncorrelated thermionic emissions. These events account for over $\SI{97}{\percent}$ of all DCs at room temperature, and their contribution drops below $\SI{70}{\percent}$ in cold xenon gas.
Rare coincident few-PE events in both PMTs suggest a mix of accidental coincidences and causal correlations, as many show negligible time delays. 

A distinct population with signal areas of approximately \SI{100}{PE} in both PMTs, visible in panel~(b), is characterized by strong time coincidence and stable rates across temperature regimes. Its magnitude and symmetry suggest an external origin, most likely cosmic muons traversing both PMT windows. These events dominate above \SI{30}{PE} (orange dashed line in panel~(b)) and account for less than \SI{1}{\percent} of all DCs at room temperature, but become more prominent in cold GXe as thermionic contributions are suppressed. 

Two additional asymmetric bands below \SI{100}{PE} are characterized by unequal sharing of signal area between the two PMTs. These are suspected to originate from internal or external radioactivity. To probe this hypothesis, a collimated $^{60}$Co source was placed at various positions outside the MarmotX cryostat. Panel~(c) shows the area correlation with the source below the PMT pair. The band corresponding to increased signal in the lower PMT (MA0058) is enhanced, confirming the influence of source positioning. When placed laterally, the asymmetry diminishes, as gamma rays are less likely to interact in both sensors. Transient increases in count rates were observed post-irradiation but returned to baseline within an hour.

Finally, when the PMTs are fully immersed in LXe, strongly correlated signals with areas that can exceed \SI{e3}{PE} emerge, as shown in panel~(d). These are attributed to xenon scintillation light induced by ambient radiation. 

\newpage

\acknowledgments
We gratefully acknowledge support from the University of Zurich, the European Research Council (ERC) under the European Union's Horizon 2020 research and innovation programme (grant agreement No. 742789, {\sl Xenoscope}), the SNF under grants 200020-219290 and 20FL20-216572, and the Dutch Research Council (NWO). We also thank Daniel Florin from the UZH Physik-Institut electronics workshop for his valuable contributions to the development  of the PMT readout base.

\newpage
\bibliographystyle{JHEP}
\bibliography{bibliography.bib}

@article{aalbersFirstDarkMatter2023,
  title = {First {{Dark Matter Search Results}} from the {{LUX-ZEPLIN}} ({{LZ}}) {{Experiment}}},
  author = {Aalbers, J. and Akerib, D. S. and Akerlof, C. W. and Al Musalhi, A. K. and Alder, F. and Alqahtani, A. and Alsum, S. K. and Amarasinghe, C. S. and Ames, A. and Anderson, T. J. and Angelides, N. and Ara{\'u}jo, H. M. and Armstrong, J. E. and Arthurs, M. and Azadi, S. and Bailey, A. J. and Baker, A. and Balajthy, J. and Balashov, S. and Bang, J. and Bargemann, J. W. and Barry, M. J. and Barthel, J. and Bauer, D. and Baxter, A. and Beattie, K. and Belle, J. and Beltrame, P. and Bensinger, J. and Benson, T. and Bernard, E. P. and Bhatti, A. and Biekert, A. and Biesiadzinski, T. P. and Birch, H. J. and Birrittella, B. and Blockinger, G. M. and Boast, K. E. and Boxer, B. and Bramante, R. and Brew, C. A. J. and Br{\'a}s, P. and Buckley, J. H. and Bugaev, V. V. and Burdin, S. and Busenitz, J. K. and Buuck, M. and Cabrita, R. and Carels, C. and Carlsmith, D. L. and Carlson, B. and {Carmona-Benitez}, M. C. and Cascella, M. and Chan, C. and Chawla, A. and Chen, H. and Cherwinka, J. J. and Chott, N. I. and Cole, A. and Coleman, J. and Converse, M. V. and Cottle, A. and Cox, G. and Craddock, W. W. and Creaner, O. and Curran, D. and Currie, A. and Cutter, J. E. and Dahl, C. E. and David, A. and Davis, J. and Davison, T. J. R. and Delgaudio, J. and Dey, S. and {de Viveiros}, L. and Dobi, A. and Dobson, J. E. Y. and Druszkiewicz, E. and Dushkin, A. and Edberg, T. K. and Edwards, W. R. and Elnimr, M. M. and Emmet, W. T. and Eriksen, S. R. and Faham, C. H. and Fan, A. and Fayer, S. and Fearon, N. M. and Fiorucci, S. and Flaecher, H. and Ford, P. and Francis, V. B. and Fraser, E. D. and Fruth, T. and Gaitskell, R. J. and Gantos, N. J. and Garcia, D. and Geffre, A. and Gehman, V. M. and Genovesi, J. and Ghag, C. and Gibbons, R. and Gibson, E. and Gilchriese, M. G. D. and Gokhale, S. and Gomber, B. and Green, J. and Greenall, A. and Greenwood, S. and {van der Grinten}, M. G. D. and Gwilliam, C. B. and Hall, C. R. and Hans, S. and Hanzel, K. and Harrison, A. and {Hartigan-O'Connor}, E. and Haselschwardt, S. J. and Hernandez, M. A. and Hertel, S. A. and Heuermann, G. and Hjemfelt, C. and Hoff, M. D. and Holtom, E. and Hor, J. Y-K. and Horn, M. and Huang, D. Q. and Hunt, D. and Ignarra, C. M. and Jacobsen, R. G. and Jahangir, O. and James, R. S. and Jeffery, S. N. and Ji, W. and Johnson, J. and Kaboth, A. C. and Kamaha, A. C. and Kamdin, K. and Kasey, V. and Kazkaz, K. and Keefner, J. and Khaitan, D. and Khaleeq, M. and Khazov, A. and Khurana, I. and Kim, Y. D. and Kocher, C. D. and Kodroff, D. and Korley, L. and Korolkova, E. V. and Kras, J. and Kraus, H. and Kravitz, S. and Krebs, H. J. and Kreczko, L. and Krikler, B. and Kudryavtsev, V. A. and Kyre, S. and Landerud, B. and Leason, E. A. and Lee, C. and Lee, J. and Leonard, D. S. and Leonard, R. and Lesko, K. T. and Levy, C. and Li, J. and Liao, F.-T. and Liao, J. and Lin, J. and Lindote, A. and Linehan, R. and Lippincott, W. H. and Liu, R. and Liu, X. and Liu, Y. and Loniewski, C. and Lopes, M. I. and Lopez Asamar, E. and L{\'o}pez Paredes, B. and Lorenzon, W. and Lucero, D. and Luitz, S. and Lyle, J. M. and Majewski, P. A. and Makkinje, J. and Malling, D. C. and Manalaysay, A. and Manenti, L. and Mannino, R. L. and Marangou, N. and Marzioni, M. F. and Maupin, C. and McCarthy, M. E. and McConnell, C. T. and McKinsey, D. N. and McLaughlin, J. and Meng, Y. and Migneault, J. and Miller, E. H. and Mizrachi, E. and Mock, J. A. and Monte, A. and Monzani, M. E. and Morad, J. A. and Morales Mendoza, J. D. and Morrison, E. and Mount, B. J. and Murdy, M. and Murphy, A. {\relax St}. J. and Naim, D. and Naylor, A. and Nedlik, C. and Nehrkorn, C. and Neves, F. and Nguyen, A. and Nikoleyczik, J. A. and Nilima, A. and O'Dell, J. and O'Neill, F. G. and O'Sullivan, K. and Olcina, I. and Olevitch, M. A. and {Oliver-Mallory}, K. C. and Orpwood, J. and Pagenkopf, D. and Pal, S. and Palladino, K. J. and Palmer, J. and Pangilinan, M. and Parveen, N. and Patton, S. J. and Pease, E. K. and Penning, B. and Pereira, C. and Pereira, G. and Perry, E. and Pershing, T. and Peterson, I. B. and Piepke, A. and Podczerwinski, J. and Porzio, D. and Powell, S. and Preece, R. M. and Pushkin, K. and Qie, Y. and Ratcliff, B. N. and Reichenbacher, J. and Reichhart, L. and Rhyne, C. A. and Richards, A. and Riffard, Q. and Rischbieter, G. R. C. and Rodrigues, J. P. and Rodriguez, A. and Rose, H. J. and Rosero, R. and Rossiter, P. and Rushton, T. and Rutherford, G. and Rynders, D. and Saba, J. S. and Santone, D. and Sazzad, A. B. M. R. and Schnee, R. W. and Scovell, P. R. and Seymour, D. and Shaw, S. and Shutt, T. and Silk, J. J. and Silva, C. and Sinev, G. and Skarpaas, K. and Skulski, W. and Smith, R. and Solmaz, M. and Solovov, V. N. and Sorensen, P. and Soria, J. and Stancu, I. and Stark, M. R. and Stevens, A. and Stiegler, T. M. and Stifter, K. and Studley, R. and Suerfu, B. and Sumner, T. J. and Sutcliffe, P. and Swanson, N. and Szydagis, M. and Tan, M. and Taylor, D. J. and Taylor, R. and Taylor, W. C. and Temples, D. J. and Tennyson, B. P. and Terman, P. A. and Thomas, K. J. and Tiedt, D. R. and Timalsina, M. and To, W. H. and Tom{\'a}s, A. and Tong, Z. and Tovey, D. R. and Tranter, J. and Trask, M. and Tripathi, M. and Tronstad, D. R. and Tull, C. E. and Turner, W. and Tvrznikova, L. and Utku, U. and Va'vra, J. and Vacheret, A. and Vaitkus, A. C. and Verbus, J. R. and Voirin, E. and Waldron, W. L. and Wang, A. and Wang, B. and Wang, J. J. and Wang, W. and Wang, Y. and Watson, J. R. and Webb, R. C. and White, A. and White, D. T. and White, J. T. and White, R. G. and Whitis, T. J. and Williams, M. and Wisniewski, W. J. and Witherell, M. S. and Wolfs, F. L. H. and Wolfs, J. D. and Woodford, S. and Woodward, D. and Worm, S. D. and Wright, C. J. and Xia, Q. and Xiang, X. and Xiao, Q. and Xu, J. and Yeh, M. and Yin, J. and Young, I. and Zarzhitsky, P. and Zuckerman, A. and Zweig, E. A.},
  year = {2023},
  month = jul,
  journal = {Phys. Rev. Lett.},
  volume = {131},
  number = {4},
  pages = {041002},
  publisher = {American Physical Society},
  doi = {10.1103/PhysRevLett.131.041002},
  urldate = {2023-10-05}
}

@article{aalbersNextgenerationLiquidXenon2022,
  title = {A Next-Generation Liquid Xenon Observatory for Dark Matter and Neutrino Physics},
  author = {Aalbers, J. and AbdusSalam, S. S. and Abe, K. and Aerne, V. and Agostini, F. and Maouloud, S. Ahmed and Akerib, D. S. and Akimov, D. Y. and Akshat, J. and Musalhi, A. K. Al and Alder, F. and Alsum, S. K. and Althueser, L. and Amarasinghe, C. S. and Amaro, F. D. and Ames, A. and Anderson, T. J. and Andrieu, B. and Angelides, N. and Angelino, E. and Angevaare, J. and Antochi, V. C. and Martin, D. Ant{\'o}n and Antunovic, B. and Aprile, E. and Ara{\'u}jo, H. M. and Armstrong, J. E. and Arneodo, F. and Arthurs, M. and Asadi, P. and Baek, S. and Bai, X. and Bajpai, D. and Baker, A. and Balajthy, J. and Balashov, S. and Balzer, M. and Bandyopadhyay, A. and Bang, J. and Barberio, E. and Bargemann, J. W. and Baudis, L. and Bauer, D. and Baur, D. and Baxter, A. and Baxter, A. L. and Bazyk, M. and Beattie, K. and Behrens, J. and Bell, N. F. and Bellagamba, L. and Beltrame, P. and Benabderrahmane, M. and Bernard, E. P. and Bertone, G. F. and Bhattacharjee, P. and Bhatti, A. and Biekert, A. and Biesiadzinski, T. P. and Binau, A. R. and Biondi, R. and Biondi, Y. and Birch, H. J. and Bishara, F. and Bismark, A. and Blanco, C. and Blockinger, G. M. and Bodnia, E. and Boehm, C. and Bolozdynya, A. I. and Bolton, P. D. and Bottaro, S. and Bourgeois, C. and Boxer, B. and Br{\'a}s, P. and Breskin, A. and Breur, P. A. and Brew, C. A. J. and Brod, J. and Brookes, E. and Brown, A. and Brown, E. and Bruenner, S. and Bruno, G. and Budnik, R. and Bui, T. K. and Burdin, S. and Buse, S. and Busenitz, J. K. and Buttazzo, D. and Buuck, M. and Buzulutskov, A. and Cabrita, R. and Cai, C. and Cai, D. and Capelli, C. and Cardoso, J. M. R. and {Carmona-Benitez}, M. C. and Cascella, M. and Catena, R. and Chakraborty, S. and Chan, C. and Chang, S. and Chauvin, A. and Chawla, A. and Chen, H. and Chepel, V. and Chott, N. I. and Cichon, D. and Chavez, A. Cimental and Cimmino, B. and Clark, M. and Co, R. T. and Colijn, A. P. and Conrad, J. and Converse, M. V. and Costa, M. and Cottle, A. and Cox, G. and Creaner, O. and Garcia, J. J. Cuenca and Cussonneau, J. P. and Cutter, J. E. and Dahl, C. E. and D'Andrea, V. and David, A. and Decowski, M. P. and Dent, J. B. and Deppisch, F. F. and de Viveiros, L. and Gangi, P. Di and Giovanni, A. Di and Pede, S. Di and Dierle, J. and Diglio, S. and Dobson, J. E. Y. and Doerenkamp, M. and Douillet, D. and Drexlin, G. and Druszkiewicz, E. and Dunsky, D. and Eitel, K. and Elykov, A. and Emken, T. and Engel, R. and Eriksen, S. R. and Fairbairn, M. and Fan, A. and Fan, J. J. and Farrell, S. J. and Fayer, S. and Fearon, N. M. and Ferella, A. and Ferrari, C. and Fieguth, A. and Fieguth, A. and Fiorucci, S. and Fischer, H. and Flaecher, H. and Flierman, M. and Florek, T. and Foot, R. and Fox, P. J. and Franceschini, R. and Fraser, E. D. and Frenk, C. S. and Frohlich, S. and Fruth, T. and Fulgione, W. and Fuselli, C. and Gaemers, P. and Gaior, R. and Gaitskell, R. J. and Galloway, M. and Gao, F. and Garcia, I. Garcia and Genovesi, J. and Ghag, C. and Ghosh, S. and Gibson, E. and Gil, W. and Giovagnoli, D. and Girard, F. and {Glade-Beucke}, R. and Gl{\"u}ck, F. and Gokhale, S. and de Gouv{\^e}a, A. and Gr{\'a}f, L. and Grandi, L. and Grigat, J. and Grinstein, B. and van der Grinten, M. G. D. and Gr{\"o}ssle, R. and Guan, H. and Guida, M. and Gumbsheimer, R. and Gwilliam, C. B. and Hall, C. R. and Hall, L. J. and Hammann, R. and Han, K. and Hannen, V. and {Hansmann-Menzemer}, S. and Harata, R. and Hardin, S. P. and Hardy, E. and Hardy, C. A. and Harigaya, K. and Harnik, R. and Haselschwardt, S. J. and Hernandez, M. and Hertel, S. A. and Higuera, A. and Hils, C. and Hochrein, S. and Hoetzsch, L. and Hoferichter, M. and Hood, N. and Hooper, D. and Horn, M. and Howlett, J. and Huang, D. Q. and Huang, Y. and Hunt, D. and Iacovacci, M. and Iaquaniello, G. and Ide, R. and Ignarra, C. M. and Iloglu, G. and Itow, Y. and Jacquet, E. and Jahangir, O. and Jakob, J. and James, R. S. and Jansen, A. and Ji, W. and Ji, X. and Joerg, F. and Johnson, J. and Joy, A. and Kaboth, A. C. and Kalhor, L. and Kamaha, A. C. and Kanezaki, K. and Kar, K. and Kara, M. and Kato, N. and Kavrigin, P. and Kazama, S. and Keaveney, A. W. and Kellerer, J. and Khaitan, D. and Khazov, A. and Khundzakishvili, G. and Khurana, I. and Kilminster, B. and Kleifges, M. and Ko, P. and Kobayashi, M. and Kodroff, D. and Koltmann, G. and Kopec, A. and Kopmann, A. and Kopp, J. and Korley, L. and Kornoukhov, V. N. and Korolkova, E. V. and Kraus, H. and Krauss, L. M. and Kravitz, S. and Kreczko, L. and Kudryavtsev, V. A. and Kuger, F. and Kumar, J. and Paredes, B. L{\'o}pez and LaCascio, L. and Laha, R. and Laine, Q. and Landsman, H. and Lang, R. F. and Leason, E. A. and Lee, J. and Leonard, D. S. and Lesko, K. T. and Levinson, L. and Levy, C. and Li, I. and Li, S. C. and Li, T. and Liang, S. and Liebenthal, C. S. and Lin, J. and Lin, Q. and Lindemann, S. and Lindner, M. and Lindote, A. and Linehan, R. and Lippincott, W. H. and Liu, X. and Liu, K. and Liu, J. and Loizeau, J. and Lombardi, F. and Long, J. and Lopes, M. I. and Asamar, E. Lopez and Lorenzon, W. and Lu, C. and Luitz, S. and Ma, Y. and Machado, P. A. N. and Macolino, C. and Maeda, T. and Mahlstedt, J. and Majewski, P. A. and Manalaysay, A. and Mancuso, A. and Manenti, L. and Manfredini, A. and Mannino, R. L. and Marangou, N. and {March-Russell}, J. and Marignetti, F. and Undagoitia, T. Marrod{\'a}n and Martens, K. and Martin, R. and {Martinez-Soler}, I. and Masbou, J. and Masson, D. and Masson, E. and Mastroianni, S. and Mastronardi, M. and {Matias-Lopes}, J. A. and McCarthy, M. E. and McFadden, N. and McGinness, E. and McKinsey, D. N. and McLaughlin, J. and McMichael, K. and Meinhardt, P. and Men{\'e}ndez, J. and Meng, Y. and Messina, M. and Midha, R. and Milisavljevic, D. and Miller, E. H. and Milosevic, B. and Milutinovic, S. and Mitra, S. A. and Miuchi, K. and Mizrachi, E. and Mizukoshi, K. and Molinario, A. and Monte, A. and Monteiro, C. M. B. and Monzani, M. E. and Moore, J. S. and Mor{\aa}, K. and Morad, J. A. and Mendoza, J. D. Morales and Moriyama, S. and Morrison, E. and Morteau, E. and Mosbacher, Y. and Mount, B. J. and Mueller, J. and Murphy, A. St J. and Murra, M. and Naim, D. and Nakamura, S. and Nash, E. and Navaieelavasani, N. and Naylor, A. and Nedlik, C. and Nelson, H. N. and Neves, F. and Newstead, J. L. and Ni, K. and Nikoleyczik, J. A. and Niro, V. and Oberlack, U. G. and Obradovic, M. and Odgers, K. and O'Hare, C. A. J. and Oikonomou, P. and Olcina, I. and {Oliver-Mallory}, K. and Oranday, A. and Orpwood, J. and Ostrovskiy, I. and Ozaki, K. and Paetsch, B. and Pal, S. and Palacio, J. and Palladino, K. J. and Palmer, J. and Panci, P. and Pandurovic, M. and Parlati, A. and Parveen, N. and Patton, S. J. and P{\v e}{\v c}, V. and Pellegrini, Q. and Penning, B. and Pereira, G. and Peres, R. and {Perez-Gonzalez}, Y. and Perry, E. and Pershing, T. and {Petrossian-Byrne}, R. and Pienaar, J. and Piepke, A. and Pieramico, G. and Pierre, M. and Piotter, M. and Pizzella, V. and Plante, G. and Pollmann, T. and Porzio, D. and Qi, J. and Qie, Y. and Qin, J. and Quevedo, F. and Raj, N. and Silva, M. Rajado and Ramanathan, K. and Garc{\'i}a, D. Ram{\'i}rez and Ravanis, J. and {Redard-Jacot}, L. and Redigolo, D. and Reichard, S. and Reichenbacher, J. and Rhyne, C. A. and Richards, A. and Riffard, Q. and Rischbieter, G. R. C. and Rocchetti, A. and Rosenfeld, S. L. and Rosero, R. and Rupp, N. and Rushton, T. and Saha, S. and Salucci, P. and Sanchez, L. and {Sanchez-Lucas}, P. and Santone, D. and dos Santos, J. M. F. and Sarnoff, I. and Sartorelli, G. and Sazzad, A. B. M. R. and Scheibelhut, M. and Schnee, R. W. and Schrank, M. and Schreiner, J. and Schulte, P. and Schulte, D. and Eissing, H. Schulze and Schumann, M. and Schwemberger, T. and Schwenk, A. and Schwetz, T. and Lavina, L. Scotto and Scovell, P. R. and Sekiya, H. and Selvi, M. and Semenov, E. and Semeria, F. and Shagin, P. and Shaw, S. and Shi, S. and Shockley, E. and Shutt, T. A. and {Si-Ahmed}, R. and Silk, J. J. and Silva, C. and Silva, M. C. and Simgen, H. and {\v S}imkovic, F. and Sinev, G. and Singh, R. and Skulski, W. and Smirnov, J. and Smith, R. and Solmaz, M. and Solovov, V. N. and Sorensen, P. and Soria, J. and Sparmann, T. J. and Stancu, I. and Steidl, M. and Stevens, A. and Stifter, K. and Strigari, L. E. and Subotic, D. and Suerfu, B. and Suliga, A. M. and Sumner, T. J. and Szabo, P. and Szydagis, M. and Takeda, A. and Takeuchi, Y. and Tan, P.-L. and Taricco, C. and Taylor, W. C. and Temples, D. J. and Terliuk, A. and Terman, P. A. and Thers, D. and Thieme, K. and Th{\"u}mmler, T. and Tiedt, D. R. and Timalsina, M. and To, W. H. and Toennies, F. and Tong, Z. and Toschi, F. and Tovey, D. R. and Tranter, J. and Trask, M. and Trinchero, G. C. and Tripathi, M. and Tronstad, D. R. and Trotta, R. and Tsai, Y. D. and Tunnell, C. D. and Turner, W. G. and Ueno, R. and Urquijo, P. and Utku, U. and Vaitkus, A. and Valerius, K. and Vassilev, E. and Vecchi, S. and Velan, V. and Vetter, S. and Vincent, A. C. and Vittorio, L. and Volta, G. and von Krosigk, B. and von Piechowski, M. and Vorkapic, D. and Wagner, C. E. M. and Wang, A. M. and Wang, B. and Wang, Y. and Wang, W. and Wang, J. J. and Wang, L.-T. and Wang, M. and Wang, Y. and Watson, J. R. and Wei, Y. and Weinheimer, C. and Weisman, E. and Weiss, M. and Wenz, D. and West, S. M. and Whitis, T. J. and Williams, M. and Wilson, M. J. and Winkler, D. and Wittweg, C. and Wolf, J. and Wolf, T. and Wolfs, F. L. H. and Woodford, S. and Woodward, D. and Wright, C. J. and Wu, V. H. S. and Wu, P. and W{\"u}stling, S. and Wurm, M. and Xia, Q. and Xiang, X. and Xing, Y. and Xu, J. and Xu, Z. and Xu, D. and Yamashita, M. and Yamazaki, R. and Yan, H. and Yang, L. and Yang, Y. and Ye, J. and Yeh, M. and Young, I. and Yu, H. B. and Yu, T. T. and Yuan, L. and Zavattini, G. and Zerbo, S. and Zhang, Y. and Zhong, M. and Zhou, N. and Zhou, X. and Zhu, T. and Zhu, Y. and Zhuang, Y. and Zopounidis, J. P. and Zuber, K. and Zupan, J.},
  year = {2022},
  month = dec,
  journal = {J. Phys. G: Nucl. Part. Phys.},
  volume = {50},
  number = {1},
  pages = {013001},
  publisher = {IOP Publishing},
  issn = {0954-3899},
  doi = {10.1088/1361-6471/ac841a},
  urldate = {2023-08-25},
  langid = {english}
}

@software{aalbers_strax,
	author = {Jelle Aalbers and Joran R. Angevaare and Daniel Wenz and Chris Tunnell and Dacheng Xu and pyup.io bot and Yossi Mosbacher and Peter Gaemers and Darryl Masson and Boris Bauermeister and Daniel Coderre and tianyu zhu and Sid El Moctar AHMED MAOULOUD and Erik Hogenbirk and Aaron Higuera and Lanqing Yuan and Joe Howlett and Sophia Farrell and Melih Kara and Alexander Bismark and Carlo Fuselli and Dominick Cichon and Evan Shockley and Jianyang Qi and Shenyang Shi and Shixiao Liang},
	doi = {10.5281/zenodo.11355772},
	month = may,
	publisher = {Zenodo},
	title = {AxFoundation/strax: v1.6.4},
	url = {https://doi.org/10.5281/zenodo.11355772},
	version = {v1.6.4},
	year = 2024}

@article{abdukerimPandaXxTMultitentonneLiquid2024,
    author = {Abdukerim, Abdusalam and Bo, Zihao and Chen, Wei and Chen, Xun and Cheng, Chen and Cheng, Zhaokan and Cui, Xiangyi and Fan, Yingjie and Fang, Deqing and Geng, Lisheng and Giboni, Karl and Gu, Linhui and Guo, Xunan and Guo, Xuyuan and Guo, Zhichao and Han, Chencheng and Han, Ke and He, Changda and He, Jinrong and Huang, Di and Huang, Junting and Huang, Zhou and Hou, Ruquan and Hou, Yu and Ji, Xiangdong and Ju, Yonglin and Li, Chenxiang and Li, Jiafu and Li, Mingchuan and Li, Shuaijie and Li, Tao and Lin, Qing and Liu, Jianglai and Lu, Congcong and Lu, Xiaoying and Luo, Lingyin and Luo, Yunyang and Ma, Wenbo and Ma, Yugang and Mao, Yajun and Meng, Yue and Ning, Xuyang and Pang, Binyu and Qi, Ningchun and Qian, Zhicheng and Ren, Xiangxiang and Shaheed, Nasir and Shang, Xiaofeng and Shao, Xiyuan and Shen, Guofang and Si, Lin and Sun, Wenliang and Tao, Yi and Wang, Anqing and Wang, Meng and Wang, Qiuhong and Wang, Shaobo and Wang, Siguang and Wang, Wei and Wang, Xiuli and Wang, Xu and Wang, Zhou and Wei, Yuehuan and Wu, Mengmeng and Wu, Weihao and Wu, Yuan and Xiao, Mengjiao and Xiao, Xiang and Yan, Binbin and Yan, Xiyu and Yang, Yong and Yu, Chunxu and Yuan, Ying and Yuan, Zhe and Yun, Youhui and Zeng, Xinning and Zhang, Minzhen and Zhang, Peng and Zhang, Shibo and Zhang, Shu and Zhang, Tao and Zhang, Wei and Zhang, Yang and Zhang, Yingxin and Zhang, Yuanyuan and Zhao, Li and Zhou, Jifang and Zhou, Ning and Zhou, Xiaopeng and Zhou, Yong and Zhou, Yubo and Zhou, Zhizhen},
    title = "{PandaX-xT\textemdash{}A deep underground multi-ten-tonne liquid xenon observatory}",
    doi = "10.1007/s11433-024-2539-y",
    journal = "Sci. China Phys. Mech. Astron.",
    volume = "68",
    number = "2",
    pages = "221011",
    year = "2025"
}

@mastersthesis{adroverCharacterizationNovelHamamatsu2023,
  title = {Characterization of the {{Novel Hamamatsu R12699-406-M4}} 2'' {{Photomultiplier Tube}} in {{Xenon}} for the {{DARWIN Experiment}}},
  author = {Adrover, M.},
  year = {2023},
  address = {Zurich, Switzerland},
  school = {University of Zurich}
}

@article{agostiniSensitivityDARWINObservatory2020f,
  title = {Sensitivity of the {{DARWIN}} Observatory to the Neutrinoless Double Beta Decay of $^{136}${{Xe}}},
  author = {Agostini, F. and Maouloud, S. E. M. Ahmed and Althueser, L. and Amaro, F. and Antunovic, B. and Aprile, E. and Baudis, L. and Baur, D. and Biondi, Y. and Bismark, A. and Breur, P. A. and Brown, A. and Bruno, G. and Budnik, R. and Capelli, C. and Cardoso, J. and Cichon, D. and Clark, M. and Colijn, A. P. and {Cuenca-Garc{\'i}a}, J. J. and Cussonneau, J. P. and Decowski, M. P. and Depoian, A. and Dierle, J. and Gangi, P. Di and Giovanni, A. Di and Diglio, S. and dos Santos, J. M. F. and Drexlin, G. and Eitel, K. and Engel, R. and Ferella, A. D. and Fischer, H. and Galloway, M. and Gao, F. and Girard, F. and Gl{\"u}ck, F. and Grandi, L. and Gr{\"o}{\ss}le, R. and Gumbsheimer, R. and {Hansmann-Menzemer}, S. and J{\"o}rg, F. and Khundzakishvili, G. and Kopec, A. and Kuger, F. and Krauss, L. M. and Landsman, H. and Lang, R. F. and Lindemann, S. and Lindner, M. and Lopes, J. A. M. and Villalpando, A. Loya and Macolino, C. and Manfredini, A. and Undagoitia, T. Marrod{\'a}n and Masbou, J. and Masson, E. and Meinhardt, P. and Milutinovic, S. and Molinario, A. and Monteiro, C. M. B. and Murra, M. and Oberlack, U. G. and Pandurovic, M. and Peres, R. and Pienaar, J. and Pierre, M. and Pizzella, V. and Qin, J. and Garc{\'i}a, D. Ram{\'i}rez and Reichard, S. and Rupp, N. and {Sanchez-Lucas}, P. and Sartorelli, G. and Schulte, D. and Schumann, M. and Lavina, L. Scotto and Selvi, M. and Silva, M. and Simgen, H. and Steidl, M. and Terliuk, A. and Therreau, C. and Thers, D. and Thieme, K. and Trotta, R. and Tunnell, C. D. and Valerius, K. and Volta, G. and Vorkapic, D. and Weinheimer, C. and Wittweg, C. and Wolf, J. and Zopounidis, J. P. and Zuber, K. and {DARWIN Collaboration}},
  year = {2020},
  month = sep,
  journal = {Eur. Phys. J. C},
  volume = {80},
  number = {9},
  pages = {808},
  issn = {1434-6052},
  doi = {10.1140/epjc/s10052-020-8196-z},
  urldate = {2023-08-25},
  langid = {english}
}

@article{AKIMOV20151,
title = {Observation of light emission from Hamamatsu R11410-20 photomultiplier tubes},
journal = {Nuclear Instruments and Methods in Physics Research Section A: Accelerators, Spectrometers, Detectors and Associated Equipment},
volume = {794},
pages = {1-2},
year = {2015},
issn = {0168-9002},
doi = {https://doi.org/10.1016/j.nima.2015.04.066},
url = {https://www.sciencedirect.com/science/article/pii/S0168900215005823},
author = {D.Yu. Akimov and A.I. Bolozdynya and Yu.V. Efremenko and V.A. Kaplin and A.V. Khromov and Yu.A. Melikyan and V.V. Sosnovtsev}
}

@software{angevaare_amstrax,
	author = {Joran R. Angevaare and Peter Gaemers and Serena Di Pede and Carlo Fuselli and Maricke Flierman and Maurice Geijsen},
	doi = {10.5281/zenodo.10281360},
	month = dec,
	publisher = {Zenodo},
	title = {XAMS-nikhef/amstrax: Release v2.1.0},
	url = {https://doi.org/10.5281/zenodo.10281360},
	version = {v2.1.0},
	year = 2023}

@article{antochiImprovedQualityTests2021,
  title = {Improved Quality Tests of {{R11410-21}} Photomultiplier Tubes for the {{XENONnT}} Experiment},
  author = {Antochi, V. C. and Baudis, L. and Bollig, J. and Brown, A. and Budnik, R. and Cichon, D. and Conrad, J. and Ferella, A. D. and Galloway, M. and Hoetzsch, L. and Kazama, S. and Koltman, G. and Landsman, H. and Lindner, M. and Mahlstedt, J. and Undagoitia, T. Marrod{\'a}n and Pelssers, B. and Volta, G. and Wack, O. and Wulf, J.},
  year = {2021},
  month = aug,
  journal = {J. Inst.},
  volume = {16},
  number = {08},
  pages = {P08033},
  publisher = {IOP Publishing},
  issn = {1748-0221},
  doi = {10.1088/1748-0221/16/08/P08033},
  urldate = {2023-11-01},
  langid = {english}
}

@article{aprileDesignPerformanceXENON102011,
  title = {Design and Performance of the {{XENON10}} Dark Matter Experiment},
  author = {Aprile, E. and Angle, J. and Arneodo, F. and Baudis, L. and Bernstein, A. and Bolozdynya, A. and Brusov, P. and Coelho, L. C. C. and Dahl, C. E. and DeViveiros, L. and Ferella, A. D. and Fernandes, L. M. P. and Fiorucci, S. and Gaitskell, R. J. and Giboni, K. L. and Gomez, R. and Hasty, R. and Kastens, L. and Kwong, J. and Lopes, J. A. M. and Madden, N. and Manalaysay, A. and Manzur, A. and McKinsey, D. N. and Monzani, M. E. and Ni, K. and Oberlack, U. and Orboeck, J. and Orlandi, D. and Plante, G. and Santorelli, R. and {dos Santos}, J. M. F. and Shagin, P. and Shutt, T. and Sorensen, P. and Schulte, S. and Tatananni, E. and Winant, C. and Yamashita, M.},
  year = {2011},
  month = apr,
  journal = {Astroparticle Physics},
  volume = {34},
  number = {9},
  pages = {679--698},
  issn = {0927-6505},
  doi = {10.1016/j.astropartphys.2011.01.006},
  urldate = {2023-10-26}
}

@article{AprileDoke2010,
	author = {Aprile, E. and Doke, T.},
	doi = {10.1103/RevModPhys.82.2053},
	issn = {00346861},
	journal = {Reviews of Modern Physics},
	number = {3},
	pages = {2053--2097},
	title = {{Liquid xenon detectors for particle physics and astrophysics}},
	volume = {82},
	year = {2010}}

@article{aprileFirstDarkMatter2023,
  title = {First {{Dark Matter Search}} with {{Nuclear Recoils}} from the {{XENONnT Experiment}}},
  author = {Aprile, E. and Abe, K. and Agostini, F. and Ahmed Maouloud, S. and Althueser, L. and Andrieu, B. and Angelino, E. and Angevaare, J. R. and Antochi, V. C. and Ant{\'o}n Martin, D. and Arneodo, F. and Baudis, L. and Baxter, A. L. and Bazyk, M. and Bellagamba, L. and Biondi, R. and Bismark, A. and Brookes, E. J. and Brown, A. and Bruenner, S. and Bruno, G. and Budnik, R. and Bui, T. K. and Cai, C. and Cardoso, J. M. R. and Cichon, D. and Cimental Chavez, A. P. and Colijn, A. P. and Conrad, J. and {Cuenca-Garc{\'i}a}, J. J. and Cussonneau, J. P. and D'Andrea, V. and Decowski, M. P. and Di Gangi, P. and Di Pede, S. and Diglio, S. and Eitel, K. and Elykov, A. and Farrell, S. and Ferella, A. D. and Ferrari, C. and Fischer, H. and Flierman, M. and Fulgione, W. and Fuselli, C. and Gaemers, P. and Gaior, R. and Gallo Rosso, A. and Galloway, M. and Gao, F. and {Glade-Beucke}, R. and Grandi, L. and Grigat, J. and Guan, H. and Guida, M. and Hammann, R. and Higuera, A. and Hils, C. and Hoetzsch, L. and Hood, N. F. and Howlett, J. and Iacovacci, M. and Itow, Y. and Jakob, J. and Joerg, F. and Joy, A. and Kato, N. and Kara, M. and Kavrigin, P. and Kazama, S. and Kobayashi, M. and Koltman, G. and Kopec, A. and Kuger, F. and Landsman, H. and Lang, R. F. and Levinson, L. and Li, I. and Li, S. and Liang, S. and Lindemann, S. and Lindner, M. and Liu, K. and Loizeau, J. and Lombardi, F. and Long, J. and Lopes, J. A. M. and Ma, Y. and Macolino, C. and Mahlstedt, J. and Mancuso, A. and Manenti, L. and Marignetti, F. and Marrod{\'a}n Undagoitia, T. and Martens, K. and Masbou, J. and Masson, D. and Masson, E. and Mastroianni, S. and Messina, M. and Miuchi, K. and Mizukoshi, K. and Molinario, A. and Moriyama, S. and Mor{\aa}, K. and Mosbacher, Y. and Murra, M. and M{\"u}ller, J. and Ni, K. and Oberlack, U. and Paetsch, B. and Palacio, J. and Peres, R. and Peters, C. and Pienaar, J. and Pierre, M. and Pizzella, V. and Plante, G. and Qi, J. and Qin, J. and Ram{\'i}rez Garc{\'i}a, D. and Singh, R. and Sanchez, L. and {dos Santos}, J. M. F. and Sarnoff, I. and Sartorelli, G. and Schreiner, J. and Schulte, D. and Schulte, P. and Schulze Ei{\ss}ing, H. and Schumann, M. and Scotto Lavina, L. and Selvi, M. and Semeria, F. and Shagin, P. and Shi, S. and Shockley, E. and Silva, M. and Simgen, H. and Takeda, A. and Tan, P.-L. and Terliuk, A. and Thers, D. and Toschi, F. and Trinchero, G. and Tunnell, C. and T{\"o}nnies, F. and Valerius, K. and Volta, G. and Weinheimer, C. and Weiss, M. and Wenz, D. and Wittweg, C. and Wolf, T. and Wu, V. H. S. and Xing, Y. and Xu, D. and Xu, Z. and Yamashita, M. and Yang, L. and Ye, J. and Yuan, L. and Zavattini, G. and Zhong, M. and Zhu, T.},
  year = {2023},
  month = jul,
  journal = {Phys. Rev. Lett.},
  volume = {131},
  number = {4},
  pages = {041003},
  publisher = {American Physical Society},
  doi = {10.1103/PhysRevLett.131.041003},
  urldate = {2023-08-25},
}

@article{aprileLoweringRadioactivityPhotomultiplier2015,
  title = {Lowering the Radioactivity of the Photomultiplier Tubes for the {{XENON1T}} Dark Matter Experiment},
  author = {Aprile, E. and Agostini, F. and Alfonsi, M. and Arazi, L. and Arisaka, K. and Arneodo, F. and Auger, M. and Balan, C. and Barrow, P. and Baudis, L. and Bauermeister, B. and Behrens, A. and Beltrame, P. and Brown, A. and Brown, E. and Bruenner, S. and Bruno, G. and Budnik, R. and B{\"u}tikofer, L. and Cardoso, J. M. R. and Coderre, D. and Colijn, A. P. and Contreras, H. and Cussonneau, J. P. and Decowski, M. P. and Giovanni, A. Di and Duchovni, E. and Fattori, S. and Ferella, A. D. and Fieguth, A. and Fulgione, W. and Galloway, M. and Garbini, M. and Geis, C. and Goetzke, L. W. and Grignon, C. and Gross, E. and Hampel, W. and Itay, R. and Kaether, F. and Kessler, G. and Kish, A. and Landsman, H. and Lang, R. F. and Calloch, M. Le and Lellouch, D. and Levinson, L. and Levy, C. and Lindemann, S. and Lindner, M. and Lopes, J. A. M. and Lyashenko, A. and Macmullin, S. and Undagoitia, T. Marrod{\'a}n and Masbou, J. and Massoli, F. V. and Mayani, D. and Fernandez, A. J. Melgarejo and Meng, Y. and Messina, M. and Miguez, B. and Molinario, A. and Murra, M. and Naganoma, J. and Oberlack, U. and Orrigo, S. E. A. and Pakarha, P. and Pantic, E. and Persiani, R. and Piastra, F. and Pienaar, J. and Plante, G. and Priel, N. and Rauch, L. and Reichard, S. and Reuter, C. and Rizzo, A. and Rosendahl, S. and {dos Santos}, J. M. F. and Sartorelli, G. and Schindler, S. and Schreiner, J. and Schumann, M. and Lavina, L. Scotto and Selvi, M. and Shagin, P. and Simgen, H. and Teymourian, A. and Thers, D. and Tiseni, A. and Trinchero, G. and Tunnell, C. and Vitells, O. and Wall, R. and Wang, H. and Weber, M. and Weinheimer, C. and Laubenstein, M. and {XENON Collaboration}},
  year = {2015},
  month = nov,
  journal = {Eur. Phys. J. C},
  volume = {75},
  number = {11},
  pages = {546},
  issn = {1434-6052},
  doi = {10.1140/epjc/s10052-015-3657-5},
  urldate = {2023-11-01},
  langid = {english}
}

@article{aprileMaterialRadiopurityControl2022,
  title = {Material Radiopurity Control in the {{XENONnT}} Experiment},
  author = {Aprile, E. and Abe, K. and Agostini, F. and Ahmed Maouloud, S. and Alfonsi, M. and Althueser, L. and Angelino, E. and Angevaare, J. R. and Antochi, V. C. and Ant{\'o}n Martin, D. and Arneodo, F. and Baudis, L. and Baxter, A. L. and Bellagamba, L. and Biondi, R. and Bismark, A. and Brown, A. and Bruenner, S. and Bruno, G. and Budnik, R. and Capelli, C. and Cardoso, J. M. R. and Cichon, D. and Cimmino, B. and Clark, M. and Colijn, A. P. and Conrad, J. and {Cuenca-Garc{\'i}a}, J. J. and Cussonneau, J. P. and D'Andrea, V. and Decowski, M. P. and Gangi, P. Di and Pede, S. Di and Giovanni, A. Di and Stefano, R. Di and Diglio, S. and Elykov, A. and Farrell, S. and Ferella, A. D. and Fischer, H. and Fulgione, W. and Gaemers, P. and Gaior, R. and Galloway, M. and Gao, F. and {Glade-Beucke}, R. and Grandi, L. and Grigat, J. and Higuera, A. and Hils, C. and Hiraide, K. and Hoetzsch, L. and Howlett, J. and Iacovacci, M. and Itow, Y. and Jakob, J. and Joerg, F. and Kato, N. and Kavrigin, P. and Kazama, S. and Kobayashi, M. and Koltman, G. and Kopec, A. and Landsman, H. and Lang, R. F. and Levinson, L. and Li, I. and Liang, S. and Lindemann, S. and Lindner, M. and Liu, K. and Lombardi, F. and Long, J. and Lopes, J. A. M. and Ma, Y. and Macolino, C. and Mahlstedt, J. and Mancuso, A. and Manenti, L. and Manfredini, A. and Marignetti, F. and Marrod{\'a}n Undagoitia, T. and Martens, K. and Masbou, J. and Masson, D. and Masson, E. and Mastroianni, S. and Messina, M. and Miuchi, K. and Mizukoshi, K. and Molinario, A. and Moriyama, S. and Mor{\aa}, K. and Mosbacher, Y. and Murra, M. and Ni, K. and Oberlack, U. and Palacio, J. and Peres, R. and Pienaar, J. and Pierre, M. and Pizzella, V. and Plante, G. and Qi, J. and Qin, J. and Ram{\'i}rez Garc{\'i}a, D. and Reichard, S. and Rocchetti, A. and Rupp, N. and Sanchez, L. and {dos Santos}, J. M. F. and Sartorelli, G. and Schreiner, J. and Schulte, D. and Schulze Ei{\ss}ing, H. and Schumann, M. and Lavina, L. Scotto and Selvi, M. and Semeria, F. and Shagin, P. and Shockley, E. and Silva, M. and Simgen, H. and Takeda, A. and Tan, P. L. and Terliuk, A. and Therreau, C. and Thers, D. and Toschi, F. and Trinchero, G. and Tunnell, C. and T{\"o}nnies, F. and Valerius, K. and Volta, G. and Wei, Y. and Weinheimer, C. and Weiss, M. and Wenz, D. and Westermann, J. and Wittweg, C. and Wolf, T. and Xu, Z. and Yamashita, M. and Yang, L. and Ye, J. and Yuan, L. and Zavattini, G. and Zhang, Y. and Zhong, M. and Zhu, T. and Zopounidis, J. P. and Laubenstein, M. and Nisi, S. and {XENON Collaboration}},
  year = {2022},
  month = jul,
  journal = {Eur. Phys. J. C},
  volume = {82},
  number = {7},
  pages = {599},
  issn = {1434-6052},
  doi = {10.1140/epjc/s10052-022-10345-6},
  urldate = {2023-08-25},
  langid = {english}
}

@article{aprileMaterialScreeningSelection2011,
  title = {Material Screening and Selection for {{XENON100}}},
  author = {Aprile, E. and Arisaka, K. and Arneodo, F. and Askin, A. and Baudis, L. and Behrens, A. and Bokeloh, K. and Brown, E. and Cardoso, J. M. R. and Choi, B. and Cline, D. and Fattori, S. and Ferella, A. D. and Giboni, K. L. and Kish, A. and Lam, C. W. and Lamblin, J. and Lang, R. F. and Lim, K. E. and Lopes, J. A. M. and Marrod{\'a}n Undagoitia, T. and Mei, Y. and Melgarejo Fernandez, A. J. and Ni, K. and Oberlack, U. and Orrigo, S. E. A. and Pantic, E. and Plante, G. and Ribeiro, A. C. C. and Santorelli, R. and {dos Santos}, J. M. F. and Schumann, M. and Shagin, P. and Teymourian, A. and Thers, D. and Tziaferi, E. and Wang, H. and Weinheimer, C. and Laubenstein, M. and Nisi, S.},
  year = {2011},
  month = sep,
  journal = {Astroparticle Physics},
  volume = {35},
  number = {2},
  pages = {43--49},
  issn = {0927-6505},
  doi = {10.1016/j.astropartphys.2011.06.001},
  urldate = {2023-08-29}
}

@article{aprileObservationTwoneutrinoDouble2019,
  title = {Observation of Two-Neutrino Double Electron Capture in {{124Xe}} with {{XENON1T}}},
  author = {Aprile, E. and Aalbers, J. and Agostini, F. and Alfonsi, M. and Althueser, L. and Amaro, F. D. and Anthony, M. and Antochi, V. C. and Arneodo, F. and Baudis, L. and Bauermeister, B. and Benabderrahmane, M. L. and Berger, T. and Breur, P. A. and Brown, A. and Brown, A. and Brown, E. and Bruenner, S. and Bruno, G. and Budnik, R. and Capelli, C. and Cardoso, J. M. R. and Cichon, D. and Coderre, D. and Colijn, A. P. and Conrad, J. and Cussonneau, J. P. and Decowski, M. P. and {de Perio}, P. and Di Gangi, P. and Di Giovanni, A. and Diglio, S. and Elykov, A. and Eurin, G. and Fei, J. and Ferella, A. D. and Fieguth, A. and Fulgione, W. and Rosso, A. Gallo and Galloway, M. and Gao, F. and Garbini, M. and Grandi, L. and Greene, Z. and Hasterok, C. and Hogenbirk, E. and Howlett, J. and Iacovacci, M. and Itay, R. and Joerg, F. and Kaminsky, B. and Kazama, S. and Kish, A. and Koltman, G. and Kopec, A. and Landsman, H. and Lang, R. F. and Levinson, L. and Lin, Q. and Lindemann, S. and Lindner, M. and Lombardi, F. and Lopes, J. A. M. and Fune, E. L{\'o}pez and Macolino, C. and Mahlstedt, J. and Manfredini, A. and Marignetti, F. and Undagoitia, T. Marrod{\'a}n and Masbou, J. and Masson, D. and Mastroianni, S. and Messina, M. and Micheneau, K. and Miller, K. and Molinario, A. and Mor{\aa}, K. and Murra, M. and Naganoma, J. and Ni, K. and Oberlack, U. and Odgers, K. and Pelssers, B. and Peres, R. and Piastra, F. and Pienaar, J. and Pizzella, V. and Plante, G. and Podviianiuk, R. and Priel, N. and Qiu, H. and Garc{\'i}a, D. Ram{\'i}rez and Reichard, S. and Riedel, B. and Rizzo, A. and Rocchetti, A. and Rupp, N. and {dos Santos}, J. M. F. and Sartorelli, G. and {\v S}ar{\v c}evi{\'c}, N. and Scheibelhut, M. and Schindler, S. and Schreiner, J. and Schulte, D. and Schumann, M. and Lavina, L. Scotto and Selvi, M. and Shagin, P. and Shockley, E. and Silva, M. and Simgen, H. and Therreau, C. and Thers, D. and Toschi, F. and Trinchero, G. and Tunnell, C. and Upole, N. and Vargas, M. and Wack, O. and Wang, H. and Wang, Z. and Wei, Y. and Weinheimer, C. and Wenz, D. and Wittweg, C. and Wulf, J. and Ye, J. and Zhang, Y. and Zhu, T. and Zopounidis, J. P. and {XENON Collaboration*}},
  year = {2019},
  month = apr,
  journal = {Nature},
  volume = {568},
  number = {7753},
  pages = {532--535},
  publisher = {Nature Publishing Group},
  issn = {1476-4687},
  doi = {10.1038/s41586-019-1124-4},
  urldate = {2023-10-20},
  copyright = {2019 The Author(s), under exclusive licence to Springer Nature Limited},
  langid = {english}
}

@article{aprileXENONnTDarkMatter2024a,
  title = {The {{XENONnT Dark Matter Experiment}}},
  author = {Aprile, E. and Aalbers, J. and Abe, K. and Maouloud, S. Ahmed and Althueser, L. and Andrieu, B. and Angelino, E. and Angevaare, J. R. and Antochi, V. C. and Martin, D. Ant{\'o}n and Arneodo, F. and Balata, M. and Baudis, L. and Baxter, A. L. and Bazyk, M. and Bellagamba, L. and Biondi, R. and Bismark, A. and Brookes, E. J. and Brown, A. and Bruenner, S. and Bruno, G. and Budnik, R. and Bui, T. K. and Cai, C. and Cardoso, J. M. R. and Cassese, F. and Chiarini, A. and Cichon, D. and Chavez, A. P. Cimental and Colijn, A. P. and Conrad, J. and Corrieri, R. and {Cuenca-Garc{\'i}a}, J. J. and Cussonneau, J. P. and Dadoun, O. and D'Andrea, V. and Decowski, M. P. and De Fazio, B. and Di Gangi, P. and Diglio, S. and Disdier, J. M. and Douillet, D. and Eitel, K. and Elykov, A. and Farrell, S. and Ferella, A. D. and Ferrari, C. and Fischer, H. and Flierman, M. and Form, S. and Front, D. and Fulgione, W. and Fuselli, C. and Gaemers, P. and Gaior, R. and Rosso, A. Gallo and Galloway, M. and Gao, F. and Gardner, R. and Garroum, N. and {Glade-Beucke}, R. and Grandi, L. and Grigat, J. and Guan, H. and Guerzoni, M. and Guida, M. and Hammann, R. and Higuera, A. and Hils, C. and Hoetzsch, L. and Hood, N. F. and Howlett, J. and Huhmann, C. and Iacovacci, M. and Iaquaniello, G. and Iven, L. and Itow, Y. and Jakob, J. and Joerg, F. and Joy, A. and Kara, M. and Kavrigin, P. and Kazama, S. and Kobayashi, M. and Koltman, G. and Kopec, A. and Kuger, F. and Landsman, H. and Lang, R. F. and Levinson, L. and Li, I. and Li, S. and Liang, S. and Lindemann, S. and Lindner, M. and Liu, K. and Loizeau, J. and Lombardi, F. and Long, J. and Lopes, J. A. M. and Ma, Y. and Macolino, C. and Mahlstedt, J. and Mancuso, A. and Manenti, L. and Marignetti, F. and Undagoitia, T. Marrod{\'a}n and Martella, P. and Martens, K. and Masbou, J. and Masson, D. and Masson, E. and Mastroianni, S. and Mele, E. and Messina, M. and Michinelli, R. and Miuchi, K. and Molinario, A. and Moriyama, S. and Mor{\aa}, K. and Mosbacher, Y. and Murra, M. and M{\"u}ller, J. and Ni, K. and Nisi, S. and Oberlack, U. and Orlandi, D. and Othegraven, R. and Paetsch, B. and Palacio, J. and Parlati, S. and Paschos, P. and Pellegrini, Q. and Peres, R. and Peters, C. and Pienaar, J. and Pierre, M. and Plante, G. and Pollmann, T. R. and Qi, J. and Qin, J. and Garc{\'i}a, D. Ram{\'i}rez and Rynge, M. and Shi, J. and Singh, R. and Sanchez, L. and dos Santos, J. M. F. and Sarnoff, I. and Sartorelli, G. and Schreiner, J. and Schulte, D. and Schulte, P. and Ei{\ss}ing, H. Schulze and Schumann, M. and Lavina, L. Scotto and Selvi, M. and Semeria, F. and Shagin, P. and Shi, S. and Shockley, E. and Silva, M. and Simgen, H. and Stephen, J. and Stern, M. and Stillwell, B. K. and Takeda, A. and Tan, P.-L. and Tatananni, D. and Terliuk, A. and Thers, D. and Toschi, F. and Trinchero, G. and Tunnell, C. and T{\"o}nnies, F. and Valerius, K. and Volta, G. and Weinheimer, C. and Weiss, M. and Wenz, D. and Westermann, J. and Wittweg, C. and Wolf, T. and Wu, V. H. S. and Xing, Y. and Xu, D. and Xu, Z. and Yamashita, M. and Yang, L. and Ye, J. and Yuan, L. and Zavattini, G. and Zhong, M. and Zhu, T.},
  year = {2024},
  month = aug,
  journal = {Eur. Phys. J. C},
  volume = {84},
  number = {8},
  pages = {784},
  issn = {1434-6052},
  doi = {10.1140/epjc/s10052-024-12982-5}
}

@article{aprileXENON100DarkMatter2012,
  title = {The {{XENON100}} Dark Matter Experiment},
  author = {Aprile, E. and Arisaka, K. and Arneodo, F. and Askin, A. and Baudis, L. and Behrens, A. and Brown, E. and Cardoso, J. M. R. and Choi, B. and Cline, D. and Fattori, S. and Ferella, A. D. and Giboni, K. L. and Kish, A. and Lam, C. W. and Lang, R. F. and Lim, K. E. and Lopes, J. A. M. and Marrod{\'a}n Undagoitia, T. and Mei, Y. and Melgarejo Fernandez, A. J. and Ni, K. and Oberlack, U. and Orrigo, S. E. A. and Pantic, E. and Plante, G. and Ribeiro, A. C. C. and Santorelli, R. and {dos Santos}, J. M. F. and Schumann, M. and Shagin, P. and Teymourian, A. and Tziaferi, E. and Wang, H. and Yamashita, M.},
  year = {2012},
  month = apr,
  journal = {Astroparticle Physics},
  volume = {35},
  number = {9},
  pages = {573--590},
  issn = {0927-6505},
  doi = {10.1016/j.astropartphys.2012.01.003},
  urldate = {2023-10-26}
}

@article{araujoUpgradedLowbackgroundGermanium2022c,
  title = {The Upgraded Low-Background Germanium Counting Facility {{Gator}} for High-Sensitivity {$\gamma$}-Ray Spectrometry},
  author = {Araujo, Gabriela R. and Baudis, Laura and Biondi, Yanina and Bismark, Alexander and Galloway, Michelle},
  year = {2022},
  month = aug,
  journal = {J. Inst.},
  volume = {17},
  number = {08},
  pages = {P08010},
  publisher = {IOP Publishing},
  issn = {1748-0221},
  doi = {10.1088/1748-0221/17/08/P08010},
  urldate = {2023-08-25},
  langid = {english}
}

@article{barrowQualificationTestsR11410212017,
  title = {Qualification Tests of the {{R11410-21}} Photomultiplier Tubes for the {{XENON1T}} Detector},
  author = {Barrow, P. and Baudis, L. and Cichon, D. and Danisch, M. and Franco, D. and Kaether, F. and Kish, A. and Lindner, M. and Undagoitia, T. Marrod{\'a}n and Mayani, D. and Rauch, L. and Wei, Y. and Wulf, J.},
  year = {2017},
  month = jan,
  journal = {J. Inst.},
  volume = {12},
  number = {01},
  pages = {P01024},
  issn = {1748-0221},
  doi = {10.1088/1748-0221/12/01/P01024},
  urldate = {2023-08-29},
  langid = {english}
}

@article{Baudis:2023pzu,
    author = "Baudis, Laura",
    title = "{Dual-phase xenon time projection chambers for~rare-event searches}",
    eprint = "2311.05320",
    archivePrefix = "arXiv",
    primaryClass = "physics.ins-det",
    doi = "10.1098/rsta.2023.0083",
    journal = "Phil. Trans. Roy. Soc. Lond. A",
    volume = "382",
    number = "2266",
    pages = "20230083",
    year = "2023"
}

@article{Baudis_2013,
   title={Performance of the Hamamatsu R11410 photomultiplier tube in cryogenic xenon environments},
   volume={8},
   ISSN={1748-0221},
   url={http://dx.doi.org/10.1088/1748-0221/8/04/P04026},
   DOI={10.1088/1748-0221/8/04/p04026},
   number={04},
   journal={Journal of Instrumentation},
   publisher={IOP Publishing},
   author={Baudis, L and Behrens, A and Ferella, A and Kish, A and Undagoitia, T Marrodán and Mayani, D and Schumann, M},
   year={2013},
   month=apr, pages={P04026–P04026} }

@misc{bismarkPMTAnalysis2023,
  title = {{{PMT Analysis}}},
  author = {Bismark, Alexander},
  year = {2023},
  urldate = {2023-12-14},
  howpublished = {\url{www.github.com/Physik-Institut-UZH/PMT_Analysis}}
}

@phdthesis{bismarkTestsFundamentalsQuantum2024,
          school = {University of Zurich},
           title = {Tests of the Fundamentals of Quantum Mechanics with Low-Background Experiments and Detector Technologies for Liquid Xenon Time Projection Chambers},
          author = {Alexander Bismark},
         address = {Zurich, Switzerland},
            year = {2024},
             url = {https://doi.org/10.5167/uzh-264330}
}

@article{diwanStatisticsChargeSpectrum2020,
  title = {Statistics of the Charge Spectrum of Photo-Multipliers and Methods for Absolute Calibration},
  author = {Diwan, M. V.},
  year = {2020},
  month = feb,
  journal = {J. Inst.},
  volume = {15},
  number = {02},
  pages = {P02001},
  issn = {1748-0221},
  doi = {10.1088/1748-0221/15/02/P02001},
  urldate = {2024-02-14},
  langid = {english}
}

@article{FUJII2015293,
	author = {Keiko Fujii and Yuya Endo and Yui Torigoe and Shogo Nakamura and Tomiyoshi Haruyama and Katsuyu Kasami and Satoshi Mihara and Kiwamu Saito and Shinichi Sasaki and Hiroko Tawara},
	doi = {https://doi.org/10.1016/j.nima.2015.05.065},
	issn = {0168-9002},
	journal = {Nucl. Instrum. Meth. A},
	pages = {293-297},
	title = {High-accuracy measurement of the emission spectrum of liquid xenon in the vacuum ultraviolet region},
	url = {https://www.sciencedirect.com/science/article/pii/S016890021500724X},
	volume = {795},
	year = {2015}}

@misc{hamamatsuphotonicsk.k.FlatPanelType2020,
  title = {Flat {{Panel Type Multianode Photomultiplier Tube R12699-406-M4}}},
  author = {{Hamamatsu Photonics K.K.}},
  year = {2020},
  urldate = {2022-02-23}
}

@book{hamamatsuphotonicsk.k.PhotomultiplierTubesBasics2017,
  title = {Photomultiplier {{Tubes}}: {{Basics}} and {{Applications}}},
  author = {{Hamamatsu Photonics K.K.}},
  year = {2017},
  edition = {4},
  publisher = {{Hamamatsu Photonics K.K.}},
  urldate = {2023-11-02}
}

@misc{hamamatsuphotonicsk.k.PhotomultiplierTubeR11410212014,
  title = {Photomultiplier Tube {{R11410-21}} for {{XENON-1T}} ({{Rev}}.-3) - {{Tentative}} Data Sheet},
  author = {{Hamamatsu Photonics K.K.}},
  year = {Revised on Sep. 30 2014}
}

@article{HOGENBIRK201687,
    author = "Hogenbirk, E. and Aalbers, J. and Bader, M. and Breur, P. A. and Brown, A. and Decowski, M. P. and Tunnell, C. and Walet, R. and Colijn, A. P.",
    title = "{Commissioning of a dual-phase xenon TPC at Nikhef}",
    doi = "10.1016/j.nima.2016.09.052",
    journal = "Nucl. Instrum. Meth. A",
    volume = "840",
    pages = "87--96",
    year = "2016"
}

@phdthesis{JulienWulf,
    author = {Julien Wulf},
    title = {Direct Dark Matter Search with XENON1T and Developments for Multi-Ton Liquid Xenon Detectors},
    school = {University of Zurich} ,
    year = {2018}
}

@article{LOPEZPAREDES201856,
title = {Response of photomultiplier tubes to xenon scintillation light},
journal = {Astroparticle Physics},
volume = {102},
pages = {56-66},
year = {2018},
issn = {0927-6505},
doi = {https://doi.org/10.1016/j.astropartphys.2018.04.006},
url = {https://www.sciencedirect.com/science/article/pii/S0927650518300173},
author = {B. {López Paredes} and H.M. Araújo and F. Froborg and N. Marangou and I. Olcina and T.J. Sumner and R. Taylor and A. Tomás and A. Vacheret},
}

@article{LUNG201232,
title = {Characterization of the Hamamatsu {R}11410-10 3-{i}n. photomultiplier tube for liquid xenon dark matter direct detection experiments},
journal = {Nuclear Instruments and Methods in Physics Research Section A: Accelerators, Spectrometers, Detectors and Associated Equipment},
volume = {696},
pages = {32-39},
year = {2012},
issn = {0168-9002},
doi = {https://doi.org/10.1016/j.nima.2012.08.052},
url = {https://www.sciencedirect.com/science/article/pii/S0168900212009230},
author = {K. Lung and K. Arisaka and A. Bargetzi and P. Beltrame and A. Cahill and T. Genma and C. Ghag and D. Gordon and J. Sainz and A. Teymourian and Y. Yoshizawa},
}

@article{LZ:2025hud,
    author = "Aalbers, J. and others",
    collaboration = "LZ",
    title = "{Measurements and models of enhanced recombination following inner-shell vacancies in liquid xenon}",
    eprint = "2503.05679",
    archivePrefix = "arXiv",
    primaryClass = "hep-ex",
    reportNumber = "FERMILAB-PUB-25-0185-V",
    doi = "10.1103/447w-94h3",
    journal = "Phys. Rev. D",
    volume = "112",
    number = "1",
    pages = "012024",
    year = "2025"
}

@phdthesis{mayaniparasPhotomultiplierTubesXENON1T2017,
  title = {Photomultiplier {{Tubes}} for the {{XENON1T Dark Matter Experiment}} and {{Studies}} on the {{XENON100 Electromagnetic Background}}},
  author = {Mayani Par{\`a}s, Daniel},
  year = {2017},
  address = {Zurich, Switzerland},
  school = {University of Zurich}
}

@article{PandaX:2024qfu,
    author = {Bo, Zihao and Chen, Wei and Chen, Xun and Chen, Yunhua and Cheng, Zhaokan and Cui, Xiangyi and Fan, Yingjie and Fang, Deqing and Gao, Zhixing and Geng, Lisheng and Giboni, Karl and Guo, Xunan and Guo, Xuyuan and Guo, Zichao and Han, Chencheng and Han, Ke and He, Changda and He, Jinrong and Huang, Di and Huang, Houqi and Huang, Junting and Hou, Ruquan and Hou, Yu and Ji, Xiangdong and Ji, Xiangpan and Ju, Yonglin and Li, Chenxiang and Li, Jiafu and Li, Mingchuan and Li, Shuaijie and Li, Tao and Li, Zhiyuan and Lin, Qing and Liu, Jianglai and Lu, Congcong and Lu, Xiaoying and Luo, Lingyin and Luo, Yunyang and Ma, Wenbo and Ma, Yugang and Mao, Yajun and Meng, Yue and Ning, Xuyang and Pang, Binyu and Qi, Ningchun and Qian, Zhicheng and Ren, Xiangxiang and Shan, Dong and Shang, Xiaofeng and Shao, Xiyuan and Shen, Guofang and Shen, Manbin and Sun, Wenliang and Tao, Yi and Wang, Anqing and Wang, Guanbo and Wang, Hao and Wang, Jiamin and Wang, Lei and Wang, Meng and Wang, Qiuhong and Wang, Shaobo and Wang, Siguang and Wang, Wei and Wang, Xiuli and Wang, Xu and Wang, Zhou and Wei, Yuehuan and Wu, Weihao and Wu, Yuan and Xiao, Mengjiao and Xiao, Xiang and Xiong, Kaizhi and Xu, Yifan and Yao, Shunyu and Yan, Binbin and Yan, Xiyu and Yang, Yong and Ye, Peihua and Yu, Chunxu and Yuan, Ying and Yuan, Zhe and Yun, Youhui and Zeng, Xinning and Zhang, Minzhen and Zhang, Peng and Zhang, Shibo and Zhang, Shu and Zhang, Tao and Zhang, Wei and Zhang, Yang and Zhang, Yingxin and Zhang, Yuanyuan and Zhao, Li and Zhou, Jifang and Zhou, Jiaxu and Zhou, Jiayi and Zhou, Ning and Zhou, Xiaopeng and Zhou, Yubo and Zhou, Zhizhen},
    collaboration = "PandaX",
    title = "{Dark Matter Search Results from 1.54\,\,Tonne\textperiodcentered{}Year Exposure of PandaX-4T}",
    doi = "10.1103/PhysRevLett.134.011805",
    journal = "Phys. Rev. Lett.",
    volume = "134",
    number = "1",
    pages = "011805",
    year = "2025"
}

@article{PandaX-4T:2024fls,
    author = "Bo, Zihao and others",
    collaboration = "PandaX-4T, PandaX",
    title = "{Measurement of two-neutrino double electron capture half-life of $^{124}$Xe with PandaX-4T}",
    eprint = "2411.14355",
    archivePrefix = "arXiv",
    primaryClass = "nucl-ex",
    doi = "10.1007/JHEP05(2025)119",
    journal = "JHEP",
    volume = "05",
    pages = "119",
    year = "2025"
}

@article{ParticleDataGroup:2024cfk,
    author = "Navas, S. and others",
    collaboration = "Particle Data Group",
    title = "{Review of particle physics}",
    doi = "10.1103/PhysRevD.110.030001",
    journal = "Phys. Rev. D",
    volume = "110",
    number = "3",
    pages = "030001",
    year = "2024"
}

@phdthesis{piastraMaterialsRadioassayXENON1T2017,
  title = {Materials {{Radioassay}} for the {{XENON1T Dark Matter Experiment}}, and {{Development}} of a {{Time Projection Chamber}} for the {{Study}} of {{Low-energy Nuclear Recoils}} in {{Liquid Xenon}}},
  author = {Piastra, Francesco},
  year = {2017},
  address = {Zurich, Switzerland},
  school = {University of Zurich}
}

@article{saldanhaModelIndependentApproach2017,
  title = {Model Independent Approach to the Single Photoelectron Calibration of Photomultiplier Tubes},
  author = {Saldanha, R. and Grandi, L. and Guardincerri, Y. and Wester, T.},
  year = {2017},
  month = aug,
  journal = {Nucl. Instrum. Meth. A},
  volume = {863},
  pages = {35--46},
  issn = {0168-9002},
  doi = {10.1016/j.nima.2017.02.086},
  urldate = {2023-11-05}
}

@article{xenoncollaboration2024xenonntanalysissignalreconstruction,
title = {{XENONnT} analysis: Signal reconstruction, calibration, and event selection},
  author = {Aprile, E. and Aalbers, J. and Abe, K. and Ahmed Maouloud, S. and Althueser, L. and Andrieu, B. and Angelino, E. and Angevaare, J. R. and Ant\'on Martin, D. and Arneodo, F. and Baudis, L. and Bazyk, M. and Bellagamba, L. and Biondi, R. and Bismark, A. and Boese, K. and Brown, A. and Bruno, G. and Budnik, R. and Cardoso, J. M. R. and Cimental Ch\'avez, A. P. and Colijn, A. P. and Conrad, J. and Cuenca-Garc\'{\i}a, J. J. and D'Andrea, V. and Daniel Garcia, L. C. and Decowski, M. P. and Deisting, A. and Di Donato, C. and Di Gangi, P. and Diglio, S. and Eitel, K. and Elykov, A. and Ferella, A. D. and Ferrari, C. and Fischer, H. and Flehmke, T. and Flierman, M. and Fulgione, W. and Fuselli, C. and Gaemers, P. and Gaior, R. and Galloway, M. and Gao, F. and Ghosh, S. and Giacomobono, R. and Glade-Beucke, R. and Grandi, L. and Grigat, J. and Guan, H. and Guida, M. and Gyorgy, P. and Hammann, R. and Higuera, A. and Hils, C. and Hoetzsch, L. and Hood, N. F. and Iacovacci, M. and Itow, Y. and Jakob, J. and Joerg, F. and Kaminaga, Y. and Kara, M. and Kavrigin, P. and Kazama, S. and Kobayashi, M. and Koke, D. and Kopec, A. and Kuger, F. and Landsman, H. and Lang, R. F. and Levinson, L. and Li, I. and Li, S. and Liang, S. and Lin, Y.-T. and Lindemann, S. and Lindner, M. and Liu, K. and Loizeau, J. and Lombardi, F. and Long, J. and Lopes, J. A. M. and Luce, T. and Ma, Y. and Macolino, C. and Mahlstedt, J. and Mancuso, A. and Manenti, L. and Marignetti, F. and Marrod\'an Undagoitia, T. and Martens, K. and Masbou, J. and Masson, E. and Mastroianni, S. and Melchiorre, A. and Merz, J. and Messina, M. and Michael, A. and Miuchi, K. and Molinario, A. and Moriyama, S. and Mor\aa{}, K. and Mosbacher, Y. and Murra, M. and M\"uller, J. and Ni, K. and Oberlack, U. and Paetsch, B. and Pan, Y. and Pellegrini, Q. and Peres, R. and Peters, C. and Pienaar, J. and Pierre, M. and Plante, G. and Pollmann, T. R. and Principe, L. and Qi, J. and Qin, J. and Ram\'{\i}rez Garc\'{\i}a, D. and Rajado, M. and Singh, R. and Sanchez, L. and dos Santos, J. M. F. and Sarnoff, I. and Sartorelli, G. and Schreiner, J. and Schulte, D. and Schulte, P. and Schulze Ei\ss{}ing, H. and Schumann, M. and Scotto Lavina, L. and Selvi, M. and Semeria, F. and Shagin, P. and Shi, S. and Shi, J. and Silva, M. and Simgen, H. and Takeda, A. and Tan, P.-L. and Terliuk, A. and Thers, D. and Toschi, F. and Trinchero, G. and Tunnell, C. D. and T\"onnies, F. and Valerius, K. and Vecchi, S. and Vetter, S. and Villazon Solar, F. I. and Volta, G. and Weinheimer, C. and Weiss, M. and Wenz, D. and Wittweg, C. and Wu, V. H. S. and Xing, Y. and Xu, D. and Xu, Z. and Yamashita, M. and Yang, L. and Ye, J. and Yuan, L. and Zavattini, G. and Zhong, M.},
  Xcollaboration = {XENON Collaboration},
  journal = {Phys. Rev. D},
  volume = {111},
  issue = {6},
  pages = {062006},
  numpages = {27},
  year = {2025},
  month = {Mar},
  publisher = {American Physical Society},
  doi = {10.1103/PhysRevD.111.062006},
  url = {https://link.aps.org/doi/10.1103/PhysRevD.111.062006}
}

@article{Yun:2024oxp,
	author = {Youhui Yun and Zhizhen Zhou and Baoguo An and Zhixing Gao and Ke Han and Jianglai Liu and Yuanzi Liang and Yang Liu and Yue Meng and Zhicheng Qian and Xiaofeng Shang and Lin Si and Ziyan Song and Hao Wang and Mingxin Wang and Shaobo Wang and Liangyu Wu and Weihao Wu and Yuan Wu and Binbin Yan and Xiyu Yan and Zhe Yuan and Tao Zhang and Qiang Zhao and Xinning Zeng},
	doi = {https://doi.org/10.1016/j.nima.2025.170290},
	issn = {0168-9002},
	journal = {Nucl. Instrum. Meth. A},
	pages = {170290},
	title = {A novel low-background photomultiplier tube developed for xenon based detectors},
	url = {https://www.sciencedirect.com/science/article/pii/S0168900225000919},
	volume = {1073},
	year = {2025}}

\end{document}